\documentclass[lettersize,journal]{IEEEtran}
\usepackage{amsmath,amsfonts}
\usepackage{array}
\usepackage[caption=false,font=normalsize,labelfont=sf,textfont=sf]{subfig}
\usepackage{textcomp}
\usepackage{stfloats}
\usepackage{url}
\usepackage{verbatim}
\usepackage{graphicx}
\usepackage{cite}
\usepackage{caption}
\usepackage{subcaption}
\usepackage{graphicx}
\usepackage{enumitem}
\usepackage[dvipsnames]{xcolor} 

\usepackage{booktabs}
\usepackage{tabularx}
\usepackage{array}
\usepackage{makecell}
\usepackage{multirow}
\usepackage{amssymb}
\usepackage[table]{xcolor}
\usepackage{algorithm}
\usepackage{algpseudocode}

\usepackage{xcolor}
\usepackage{tikz}
\usetikzlibrary{calc}
\usetikzlibrary{arrows.meta}

\usepackage{orcidlink}

\definecolor{OIblue}{HTML}{0072B2}
\definecolor{OIorange}{HTML}{E69F00}
\definecolor{OIgreen}{HTML}{009E73}
\definecolor{OIvermillion}{HTML}{D55E00}
\definecolor{OIgray}{HTML}{666666}

\definecolor{stdZeroZeroFive}{HTML}{481D6F}
\definecolor{stdZeroOne}{HTML}{365D8D}
\definecolor{stdZeroTwo}{HTML}{21918C}
\definecolor{stdZeroFive}{HTML}{48C16E}
\definecolor{stdZeroTen}{HTML}{CAE11F}

\newcommand{\mP}{\mathcal{P}}

\definecolor{darkblue}{RGB}{0,0,169}
\definecolor{darkred}{RGB}{139,0,0}

\newif\iftbrhidden
\tbrhiddenfalse   

\usepackage{booktabs}
\usepackage{amsthm}

\theoremstyle{definition}
\newtheorem{definition}{Definition}

\begin{document}

\title{Combining Homomorphic Encryption and Differential Privacy in Federated Learning for Model Inspection and Availability}

\author{
Ceren Yıldırım\,\orcidlink{0000-0002-8029-6520}, 
Kamer Kaya\,\orcidlink{0000-0001-8678-5467},
Sinan Yıldırım\,\orcidlink{0000-0001-7980-8990}, 
and
Erkay Savaş\,\orcidlink{0000-0002-4869-5556},~\IEEEmembership{Member,~IEEE}
    
\thanks{All authors are with Sabancı University,
    Türkiye (e-mail: \{cerenyildirim, kaya, sinanyildirim, erkays\}@sabanciuniv.edu).
    Corresponding author: Ceren Yıldırım.}%
}



\maketitle

\begin{abstract}
The increasing prevalence of decentralized data has led to a growing interest in federated learning, which enables collaborative model training without clients sharing their sensitive local data. However, FL alone does not sufficiently protect sensitive training data and is generally coupled with privacy-preserving techniques, such as differential privacy and homomorphic encryption. Although powerful, these techniques address separate concerns via different mechanisms, so relying on just one might prove insufficient or impractical for addressing challenges associated with federated learning. In this work, we propose a privacy-preserving federated learning framework that combines homomorphic encryption-based training with differential privacy-based model inspection and release. We adopt a Markov chain Monte Carlo-based Bayesian privacy estimation method to estimate the privacy of our proposed framework. Our results show that this method improves both model utility and estimated privacy over the baseline method that relies solely on differential privacy for training. In our experiments with the FEMNIST dataset, by the end of training, our method reaches a test loss of $1.09$, compared to $2.37$ for the differential privacy-only approach, while providing stronger estimated privacy protection, with the estimated posterior mean of the privacy parameter $\epsilon$ of $4.32$, compared to $7.26$ for the differential privacy-only approach. We also show that intermittent model monitoring can preserve the encrypted training trajectory while, under our evaluated experimental setting, providing estimated privacy comparable to or stronger than the differential privacy-only approach.
\end{abstract}


\section{Introduction}
\label{sec:introduction}
\IEEEPARstart{I}{n} recent years, the rapid growth of data generated and stored in a distributed fashion has driven increasing interest in Federated Learning~(FL), a Machine Learning~(ML) paradigm that enables decentralized data owners to collaboratively train a model without sharing their local data. In FL, each data owner trains a model locally using their confidential data and shares the local model updates with a server. The server aggregates the updates it receives from the clients to produce a global model update, which it redistributes to the clients. This local training and aggregation process repeats over multiple rounds until the model reaches a target accuracy.

FL has also been widely recognized as a privacy-enhancing solution as it ensures each client's sensitive data remains within its own local environment. However, recent studies have shown that local and global model updates can still reveal information about confidential training data~\cite{Li_et_al_2019, Fredrikson_et_al_2015, Shokri_et_al_2017}. For this reason, cryptographic or statistical techniques such as Homomorphic Encryption~(HE) and Differential Privacy~(DP) have become common approaches for developing privacy-preserving FL systems~\cite{Gentry2009, Dwork_2006}. 

While both DP and HE are powerful tools for enhancing privacy in FL, they are not interchangeable, as their different underlying mechanisms make them better suited to address different privacy concerns. Relying solely on either method is insufficient to ensure training data confidentiality while preserving the utility of the resulting model. DP protects sensitive training data by injecting noise into model updates during FL. However, estimating the privacy of complex ML models and calibrating noise accordingly are challenging tasks. Moreover, relying on often overly conservative worst-case theoretical bounds can significantly reduce model utility, especially when strong privacy guarantees are required.

HE allows computations to be performed directly on encrypted model updates. Although it also introduces a small amount of noise during encryption, this noise differs fundamentally from the noise deliberately added by DP to achieve privacy guarantees. In HE, noise is merely a byproduct of encryption and subsequent operations, and is typically negligible in practice. This greatly reduces the utility degradation often associated with DP, and information remains protected as long as it remains encrypted. However, once sensitive information or its outputs are decrypted, they are no longer protected by HE. For instance, the final model or its inference results released in plaintext may still reveal information about the underlying training data. Therefore, additional mechanisms, such as DP, are required for privacy-preserving downstream use.

Moreover, fully encrypted training complicates monitoring the learning process. For instance, assessing model convergence, deciding when to halt the protocol, or detecting utility-harming client behavior becomes non-trivial when model updates are encrypted. Even non-malicious clients with low-quality or biased data may inadvertently degrade global model quality. Identifying such cases during training is important to ensure that the final model remains useful. However, if the entire training process is performed under HE, these checks must also be carried out over encrypted information, making it much more difficult to inspect the progression of model training privately.

In this work, we propose a privacy-preserving FL protocol that combines HE and DP to leverage the respective advantages of both techniques. We use HE to preserve model utility during federated training, while employing DP for intermittent model inspection and privacy-preserving downstream release. The entire federated training workflow operates on encrypted and non-perturbed model updates, and DP is applied only to separate copies of the global updates that are subsequently made publicly available for inspection or downstream use. 

As noted, estimating the privacy of complex ML models and determining appropriate noise levels accordingly remain challenging problems. To address this, we use a Markov chain Monte Carlo~(MCMC)-based Bayesian estimation method for DP based on Membership Inference Attacks~(MIAs), introduced in~\cite{Yildirim_et_al_2025}. We adapt this method to assess privacy in FL by introducing a new attack definition and corresponding test statistics. This lets us estimate the full posterior distribution of the privacy parameters, accounting for uncertainty in attack performance and avoiding overly confident privacy estimates.

In our evaluation, we compare three FL scenarios that represent different privacy-utility trade-offs. These include
\begin{enumerate}[leftmargin=*]
    \item a DP-only baseline scenario where noise is injected into the local model updates before each aggregation round, 
    \item a scenario where training is fully achieved using HE and DP is applied to separate copies of model updates for inspection and release of the final model,
    \item and a scenario where training is fully achieved using HE and DP is applied only to the final model before release.
\end{enumerate}

Our findings indicate that the third setting yields lower estimated privacy loss than the first while providing better utility. Furthermore, we show that intermittent monitoring (i.e., the second setting) can be achieved without perturbing or modifying the encrypted training process, while still providing lower estimated privacy loss than that of the first setting.

\subsection{Contributions}
\label{sec:contributions}
Our main contributions can be summarized as follows:

\begin{itemize}[leftmargin=*]
    \item We introduce a fully homomorphic FL framework that supports intermittent model inspection through DP, thereby preserving model utility while enabling privacy-preserving inspection during training.
    \item We introduce a new attack strategy and corresponding test statistics to empirically estimate privacy in FL settings, inspired by the method in~\cite{Yildirim_et_al_2025}. This approach enables us to estimate the model's privacy throughout training and downstream use.
    \item Using our privacy estimation technique, we demonstrate that using HE for training and applying perturbations only to the final trained model can significantly maintain predictive accuracy while offering stronger estimated privacy guarantees compared to the DP-only alternative, as presented in Fig.~\ref{fig:loss_vs_eps}. 
    \item We demonstrate that we can conduct intermittent inspections during training in our protocol while preserving the same degree of utility and achieving estimated privacy comparable to or stronger than the DP-only baseline, with privacy increasing as inspections become less frequent (see Fig.~\ref{fig:Pp_grid}).
\end{itemize}

\subsection{Outline}
\label{sec:outline}
In Section~\ref{sec:background}, we provide information on the technical background relevant to this work, including FL, HE, and DP. In Section~\ref{sec:related-work}, we review prior studies on privacy-preserving FL that utilize HE or DP, discussing their advantages and shortcomings relative to our own proposal. Section~\ref{sec:methodology} outlines our FL framework and introduces it in detail. In Section~\ref{sec:experimental-results}, we present experimental results comparing the estimated privacy and model utility of our proposed method with those of a DP-only baseline, demonstrating the advantages of our approach. Finally, in Section~\ref{sec:conclusion}, we conclude with a discussion of our main contributions and future work.

\section{Background} 
\label{sec:background}



\subsection{Federated Learning}
\label{sec:federated-learning}
Federated Learning~(FL) is a technique that enables multiple clients to collaboratively train an ML model while keeping their local data concealed from other participants. This technique ensures that sensitive training data is never directly shared with any third parties. Instead, each client trains an ML model on its local data and shares only the gradients or parameters of the local update with a central server (hereafter referred to as the server).
These shared model parameters are combined to produce a global model update. The FL process continues for multiple rounds until the global model achieves a target accuracy. 

The most widely employed technique for FL is federated averaging (FedAvg)~\cite{mcmahan2017communication}. A training cohort $\mathcal{T}$ of size $K$ begins the protocol with a randomly initialized global model. Then, each client within $\mathcal{T}$ separately computes the gradient of the loss function on its local data and updates its model parameters. A local model update is computed as
\[
\theta_c^{(r)} \leftarrow \theta_c^{(r)}-\eta \nabla F_c(\theta_c^{(r)}),
\]
where $\theta_c^{(r)}$ is client $c$'s local model parameters at round $r$, $\eta$ is the (global) learning rate, and $\nabla F_c$ is the average gradient over client $c$'s local dataset{, denoted as $\mathcal{Z}_{c}$}. This step is performed over multiple mini-batches and epochs before the final local update is sent to the server.

At each communication round $r$, the server randomly selects a subset of clients {$\mathcal{C}^{(r)} \subseteq \mathcal{T}$}, where $|\mathcal{C}^{(r)}|=K_s$, to contribute to the aggregation. Once the server receives the local updates from the clients in $\mathcal{C}^{(r)}$, it computes a weighted average of the local updates to generate the new global model parameters
\[
\theta^{(r)} \leftarrow
\sum_{c \in \mathcal{C}^{(r)}} \frac{n_{c}}
{\sum_{c'\in \mathcal{C}^{(r)}} n_{c'}} \theta_{c}^{(r)},
\]
{where $n_{c}$ denotes the size of the local dataset $\mathcal{Z}_{c}$.} This iterative local training and global aggregation process continues for several rounds until the model reaches the desired performance level.

Several other advanced FL techniques adapt this basic FedAvg approach to handle more complex, non-i.i.d. (non-independent and identically distributed) data partitions. These aim to mitigate the convergence issues and performance degradation caused by data heterogeneity, e.g., FedProx~\cite{li2020fedprox}, Scaffold~\cite{karimireddy2020scaffold}, and FedNova~\cite{wang2020tackling}. 

\subsection{Homomorphic Encryption}
Homomorphic Encryption (HE) is a cryptographic technique that allows computation on encrypted data without requiring the decryption key. The general idea can be expressed as 
$
\operatorname{Enc}(m_1) \otimes \operatorname{Enc}(m_2) = \operatorname{Enc}(m_1 \circ m_2),
$
where $\operatorname{Enc}$ represents the encryption operation, $\otimes$ denotes an operation in the encrypted domain, and $\circ$ is an operation, e.g., addition or multiplication, in the plaintext domain. HE schemes are categorized by their computational capabilities and the complexity of functions they can evaluate. Among these, Fully Homomorphic Encryption (FHE) stands out as it can support any computable function and is not restricted by the number of consecutive operations. This is achieved through a process called \textit{bootstrapping}\cite{Gentry2009}. Bootstrapping refreshes the ciphertext by homomorphically reducing the noise that accumulates with each applied homomorphic operation, thus enabling an unlimited number of operations.

Some of the most popular FHE schemes include the Cheon-Kim-Kim-Song (CKKS)~\cite{CKKS2017}, Brakerski/Fan-Vercauteren~(BFV)~\cite{FV2012}, and Fast Fully Homomorphic Encryption over the Torus
(TFHE)~\cite{TFHE2016} schemes. CKKS supports approximate arithmetic over fixed-point and complex numbers, BFV supports exact arithmetic over integers, and TFHE is optimized for gate-level Boolean logic and arithmetic over bits. Because it supports fixed-point numbers, CKKS is generally preferred for machine learning applications. The security of these schemes is based on the hardness assumption of the Ring Learning with Errors (RLWE) problem and provides semantic security (IND-CPA). Furthermore, BFV and CKKS schemes also support \textit{batching}, an encoding technique that packs multiple plaintext values into a single ciphertext. This enables Single-Instruction, Multiple-Data (SIMD) operations, significantly improving the computational throughput.

\subsection{Differential Privacy}
\label{sec:differential-privacy}
Differential Privacy (DP) is a mathematical framework that aims to prevent an individual's information from being inferred through an algorithm's output. This is generally achieved by adding noise to the computational process. The noise level is carefully calibrated to strike a balance between privacy and utility. Excessive noise can disrupt the statistical properties of the output, while minimal noise may not provide sufficient privacy. Therefore, the noise needs to be just large enough so that the output distributions from two \emph{neighboring datasets} are nearly indistinguishable. The neighborhood relation depends on the context. Typically, two datasets are neighbors if one dataset can be obtained by adding/removing (or replacing) one individual's data from the other dataset.

\begin{definition}
A randomized mechanism $\mathcal{M}: \mathcal{U}\mapsto \mathcal{V}$ is $(\epsilon, \delta)$-DP if for all neighboring $(\mathcal{D}, \mathcal{D}')$ pairs and all $V \in \mathcal{V}$,
\[
\mathbb{P}(\mathcal{M}(\mathcal{D}) \in V) \leq e^{\epsilon} \, \mathbb{P}(\mathcal{M}(\mathcal{D}') \in V) + \delta.
\]
\end{definition}

\subsubsection{Estimating DP}
\label{sec:estimating-dp}
Although DP serves as a powerful tool for safeguarding the privacy of individual data, estimating the privacy parameters $(\epsilon, \delta)$ remains challenging. This is especially true in complex scenarios that involve multiple randomized operations, such as training sophisticated ML models. In these cases, a considerable gap often exists between theoretical guarantees and the privacy actually achieved by the model. As a result, there has been growing interest in empirical estimation of these parameters.

Empirical methodologies often leverage the relationship between DP and membership inference attacks~(MIAs) to estimate privacy. An MIA can be formulated as a statistical test in which the attacker attempts to determine whether a target sample or individual was included in a target model's training data. Various studies in the literature employ either frequentist or Bayesian frameworks for privacy estimation. For instance, \cite{Jagielski_et_al_2020, Nasr_et_al_2023} use Clopper-Pearson confidence intervals, while~\cite{Zanella-Beguelin_et_al_2023, Yildirim_et_al_2025} rely on Bayesian credible intervals to provide estimates for privacy parameters.


The effectiveness of the underlying attack is critical to privacy estimation. While the likelihood ratio test~(LRT) is the most powerful attack in principle, its direct practical application is not feasible, leading MIAs to rely on various approximations. These approximations include LOSS~\cite{Yeom_et_al_2018}, a loss-based test that is effective in identifying non-membership but performs poorly in detecting membership, and LiRa~\cite{Carlini_et_al_2022}, which provides a more direct approximation of the LRT.

\section{Related Work}
\label{sec:related-work}
Several works in the literature aim to provide privacy in FL using various techniques, including differential privacy (DP), homomorphic encryption (HE), secure multi-party computation (SMPC), trusted execution environments (TEEs), and decentralization~\cite{li_threats_2025}. In this section, we focus primarily on DP and HE, as our work uses these two methods to achieve privacy.

\paragraph{Differential Privacy} There are two main types of DP in FL: Centralized Differential Privacy (CDP) and Local Differential Privacy (LDP). In CDP, the server perturbs the global model update after aggregating the local models received from the clients~\cite {geyer_differentially_2018}. Existing works that use CDP in the domain of FL include~\cite{McMahan2017LearningDP}. This approach is applicable when the server can be trusted, an assumption that does not hold in most real-world scenarios. In LDP, by contrast, each client perturbs their local information themselves before transmitting it to the server~\cite{duchi_local_2014}. Existing works that use LDP in the domain of FL include~\cite{shokri_privacy-preserving_2015, truex_ldp-fed_2020}. Although LDP protects against an untrusted server, ensuring privacy guarantees while maintaining model utility becomes a major challenge as model size increases. This is due to the privacy budget being spread over a large parameter space, which necessitates adding more noise to the model~\cite{Duchi2016MinimaxOP}.


\paragraph{Homomorphic Encryption} 
Several works utilize HE to achieve varying degrees of privacy. In~\cite{fang_privacy_2021}, an improved version of the Paillier cryptosystem is used. In this work, all clients share the same single secret/public key pair at each round. However, this method is vulnerable to collusion attacks between the server and any client, as well as infiltration attacks, due to the secret key being shared. 

In~\cite{ma_privacy_2022}, the authors use multi-key homomorphic encryption over the CKKS encryption scheme~(xMK-CKKS). In this scheme, the secret key is divided into multiple shares, with each share owned by a different client. This approach ensures that all clients must collaborate to decrypt any ciphertext, improving upon the security guarantees of~\cite{fang_privacy_2021}. While this protocol secures local model weights, the global model remains visible to the server and is not collusion-resistant when $K-1$ clients collude, where $K$ is the total number of clients involved in the protocol. Moreover, in both~\cite{fang_privacy_2021} and~\cite{ma_privacy_2022}, the clients can access the global update in plaintext, making the model susceptible to statistical attacks.

One of the most notable works in the literature in terms of privacy protection is the \textsc{Poseidon} framework~\cite{sav_poseidon_2021}. For privacy protection, this framework uses multiparty homomorphic encryption (MHE), following a key-distribution structure similar to xMK-CKKS. In this framework, training is conducted entirely homomorphically, and model updates remain encrypted at all times. After a certain number of communication rounds, the model can either be decrypted for further analysis or kept confidential for encrypted model inference. The protocol provides end-to-end model protection; however, it has some limitations:

\begin{itemize}[leftmargin=*]
    \item Keeping the model encrypted throughout the FL process makes it infeasible to inspect its utility during training. This complicates decisions about when to stop training and makes it difficult to detect anomalies during training.
    \item To utilize the trained model in downstream tasks, the authors suggest perturbing the model's output (e.g., predictions) with DP. However, this method results in some privacy loss with each query, which accumulates over time and limits the number of queries the model can effectively process.
\end{itemize}

Our proposed method is specifically designed to address these two critical limitations of the \textsc{Poseidon} protocol.

\section{Privacy-Preserving and Inspectable FL}
\label{sec:methodology}
In this section, we first outline the system settings for our FL protocol and discuss the limitations of privacy-preserving FL. Next, we introduce our protocol, explain the MCMC-based Bayesian method we use for privacy estimation, and describe our threat model, along with the construction of the attack used for privacy estimation. 

\subsection{System Setting}
\label{sec:system-setting}
Depending on the application and use case, the \emph{population} $\mathcal{P}$ in FL can consist of (a) a few large institutions (e.g., hospitals, banks) managing data from multiple individuals, or (b) a vast number of edge clients (e.g., cellphones, IoT devices) where each client represents data from a single individual. Eq.~\eqref{eq:population} shows the differences in these two settings by defining the {training dataset} $\mathcal{Z}_{c}$ of each client $c \in \mathcal{P}$. Here, $\mathcal{U}$ denotes the collection of the datasets of all individuals, while ${\mathcal{U}_{c}} \subset \mathcal{U}$ denotes the collection of datasets of individuals linked to client $c$.
\begin{equation}
\mathcal{Z}_c = \bigcup_{U \in \mathcal{U}_c} U,
\quad
\begin{cases}
|\mathcal{U}_{c}| \geq 1 \text{ and } |\mathcal{P}| \leq |\mathcal{U}|, & \text{if (a)},\\
|\mathcal{U}_{c}| = 1 \text{ and } |\mathcal{P}| = |\mathcal{U}|, & \text{if (b)}.
\end{cases}
\label{eq:population}
\end{equation}
Each client's training dataset, $\mathcal{Z}_{c}$, is typically limited in size and lacks the feature/label distribution that reflects the real-world distribution of the relevant data type. These clients come together with the common goal of training an ML model, as their local training datasets are not representative enough to build an accurate ML model on their own.

Our FL protocol adopts the definition (b) of FL, in which each client in the population is an individual possessing a unique yet limited training dataset. We expect ensuring privacy under definition (b) to be more challenging than under definition (a), since each client's contribution directly relates to the updates from the specific individuals we aim to protect. This motivates our choice of definition (b) to establish a more conservative estimate for privacy loss. 

\subsection{Limitations of Privacy-Preserving FL}
\label{sec:limitations-of-ppfl}
In \textsc{Poseidon}, {output perturbation is suggested for providing DP}. When a query is made for a selected input, inference is performed homomorphically over the encrypted model. The resulting encrypted prediction is then perturbed with DP noise, after which the noisy prediction is decrypted and released to the querier. However, the privacy loss associated with this approach accumulates with the number of queries due to the \emph{sequential composition property} of DP, limiting how many times the final model can be used privately. 

\begin{definition}[Sequential composition property of DP~\cite{mcsherry2007mechanism}]
Let $\mathcal{M}_1, \ldots, \mathcal{M}_m$ be randomized mechanisms applied to the same dataset $\mathcal{D}$, where each $\mathcal{M}_i$ satisfies $(\epsilon_i, \delta_i)$--DP. Then, the combined output $\mathcal{M}_1(\mathcal{D}), \ldots, \mathcal{M}_m(\mathcal{D})$ is $\left(\sum_{i=1}^{m}\epsilon_i, \sum_{i=1}^{m}\delta_i\right)$-DP.
\end{definition}

To mitigate this limitation, {we take the approach of \emph{model perturbation}.} We leverage encrypted training as the foundation of our privacy-preserving FL pipeline and apply calibrated DP noise {\em to the final model} rather than to each prediction output. This approach allows the final noisy model to be queried an indefinite number of times once it is released, thanks to the \emph{post-processing property} of DP.

\begin{definition}[Post-processing property of DP~\cite{dwork2014algorithmic}]
    Let a randomized mechanism $\mathcal{M}: \mathcal{U} \rightarrow \mathcal{V}$ be ($\epsilon, \delta)$-DP. For any function $f: \mathcal{V} \rightarrow \mathsf{W}$, $f \circ \mathcal{M}$ is also $(\epsilon, \delta)$-DP. 
\end{definition}

A serious challenge within the model perturbation approach is deriving tight theoretical privacy bounds in complex ML pipelines. Overly conservative theoretical bounds may lead to unnecessarily large noise, resulting in perturbed models with very low utility. To mitigate this, we propose empirical privacy auditing to estimate the model's privacy at a given noise level, and ultimately determine the noise level based on those estimates. Privacy auditing is based on designing membership inference attacks (MIA) tailored to the DP definition, measuring the performance of MIA on several instances, and estimating DP parameters from the performance results. Several methods have been proposed for this purpose \cite{Zanella-Beguelin_et_al_2023,Yildirim_et_al_2025}. In this work, we adopt the Bayesian privacy estimation algorithm in~\cite{Yildirim_et_al_2025} and extend it to this work by adapting the attack definition and test statistic to the FL setting.

Another shortcoming of executing the full training pipeline homomorphically is that it effectively turns the model updates and any intermediate values into a ``{\em{black box}}". This makes it significantly more challenging to determine when the model has reached a satisfactory level of utility and when to stop training. Furthermore, it becomes difficult to identify anomalies that could potentially harm the training process. Clients may provide low-quality, imbalanced, or biased datasets, which can negatively affect the model without them realizing it. Therefore, it is essential to conduct model inspections during training.

One way to conduct inspections is to examine the global model updates after specific rounds. By accessing plaintext global model updates, clients can run comprehensive statistical tests or evaluate the model's classification performance on a separate test set. This method avoids additional computational and communication costs associated with HE during inspections. However, making global model updates publicly available poses a risk of privacy breaches. A better alternative is to conduct the privacy-preserving inspections on perturbed copies of these model updates during specific rounds using DP. Since the noisy model is merely a temporary copy and the training continues with the non-noisy version, the added noise will not affect the accuracy of the model being trained (see Fig.~\ref{fig:three_pipelines}(b)).

\subsection{Privacy-Preserving FL using HE}
\label{sec:ppfl-using-he}
\subsubsection{Cryptographic Setup}
\label{sec:cryptographic-setup}
Our protocol relies on multiparty homomorphic encryption (MHE)~\cite{mouchet2021multiparty}. Each client $c \in \mathcal{T}$, where $|\mathcal{T}|=K$, generates a secret key share $\mathsf{sk}_c$ using the standard secret key generation procedure of CKKS. The parties then collectively generate a public key $\mathsf{pk}$ corresponding to their joint secret key shares. Encryption and homomorphic evaluation are performed using $\mathsf{pk}$, as done in the standard single-key CKKS setting. Decryption, however, requires all secret key shares; therefore, a ciphertext can be decrypted only if all $K$ parties are present.

The parties store their data locally in cleartext form. However, the global model update received from the server at each round $r$ is encrypted and denoted by $[{\theta^{(r)}}]$, where $[\cdot]$ indicates the encryption of its argument under $\mathsf{pk}$. Since the global model updates remain encrypted throughout FL, the local training must also be performed homomorphically. As a result, for some client $c \in \mathcal{T}$, each local model update is also encrypted and represented by $[\theta^{(r)}_c]$. These encrypted local model updates are sent to the server, which aggregates them homomorphically to produce the next global model update. 

\subsubsection{Encrypted Federated Learning}
\label{sec:encrypted-fl}
A training coalition $\mathcal{T}$ joins the FL protocol. Each $c \in \mathcal{T}$ possesses a local dataset of their own, $\mathcal{Z}_{c}$. The FL protocol runs for $R$ iterative rounds. At the beginning of round $r$, the server selects and notifies a subset of the training cohort uniformly at random, $\mathcal{C}^{(r)} \subseteq \mathcal{T}$, where $|\mathcal{C}^{(r)}| = K_s$. Typically, $K_{s} \ll K$ is taken to reduce communication overhead.

The server sends the encrypted global model update, $[\theta^{(r-1)}]$ from the previous round to each selected client $c \in \mathcal{C}^{(r)}$. Then, each selected party $c \in \mathcal{C}^{(r)}$ homomorphically trains a local model on their corresponding local dataset $\mathcal{Z}_{c}$, all starting from the same global model update $[{\theta}^{(r-1)}]$. The resulting local model for client $c$ is denoted by $[\theta^{(r)}_c]$. The local model updates $\theta^{(r)}_{c}$, for $c \in \mathcal{C}^{(r)}$ are homomorphically aggregated, weighted by their respective dataset sizes:
\[
[{\theta}^{(r)}] = \sum_{c\in\mathcal{C}^{(r)}} \frac{n_{c} [\theta^{(r)}_{c}]}{\sum_{c'\in \mathcal{C}^{(r)}} n_{c'}}.
\]

\subsubsection{Differentially Private Model Inspection}
If model inspection is needed at round $r$, the server creates a separate copy of the encrypted global update $[{\theta}^{(r)}]$, generates a Gaussian noise term $V^{(r)}$, and produces a noisy version of the encrypted global update
\[
[\tilde{\theta}^{(r)}] = [{\theta}^{(r)}] + V^{(r)}.
\]
The noise addition is performed homomorphically. After generating the noisy model $[\tilde{\theta}^{(r)}]$, the clients come together to decrypt it and make it available for inspection. {Critically, t}he training process continues from the non-perturbed encrypted global update $[{\theta}^{(r)}]$.

\subsubsection{Downstream Model Release}
After completing $R$ rounds of training, the model is prepared for downstream use (e.g., inference), where the model is assumed to be in cleartext form. To achieve this, we inject a Gaussian noise term $V^{(R)}$ into the final model before decrypting and publicly releasing it, similar to our approach for intermittent updates:
\[
[\tilde{\theta}^{(R)}] = [{\theta}^{(R)}] + V^{(R)}.
\]

\subsection{DP Estimation for FL}
To estimate the privacy guarantees of our proposed FL protocol, we use the MCMC-based Bayesian estimation technique introduced in~\cite{Yildirim_et_al_2025}. Our choice of this method is motivated by its ability to estimate the full posterior distribution of the privacy parameters, rather than merely relying on credible intervals. It also consolidates evidence from multiple attacks, leading to a more robust privacy estimate. Moreover, it does not assume that any single attack is the strongest attack, or that the attack is performed on the worst-case \textit{challenge base}. Instead, it parametrizes and estimates the average strength of the applied attacks and challenge bases. Finally, the method provides an efficient approach for evaluating attack performance and can accommodate statistical dependencies, as it naturally extends to joint probability models.

The estimation procedure can be represented as a hierarchical model over $(\epsilon, s)$ parameters, denoting the privacy and attack strength. Ideally, the attack's true type I (false positive) and type II (false negative) error probabilities are used as the input to the privacy estimation algorithm; however, these are latent values that can only be estimated via error counts obtained through multiple attack executions. The auditing procedure can be outlined in three main steps:
\begin{itemize}[leftmargin=*]
    \item \textit{Designing the attack:} \cite{Yildirim_et_al_2025} is based on a parametric version of LiRA~\cite{Carlini_et_al_2022}, which is a loss-based attack. For privacy estimation, the choice of the \emph{test statistic} is crucial and depends on the use case and attack definition. Since the FL setting considered in this work differs from the original setting, we introduce a new attack definition and adapt the test statistic accordingly. The resulting attack formulation is described in Section~\ref{sec:threat-model}.
    \item \textit{Measuring attack performance and obtaining error counts:} To audit the privacy of an algorithm, the attack must be repeated multiple times to estimate the type I and type II error probabilities. 
    \item \textit{Estimating privacy from the error counts:} Given the estimated error probabilities, we estimate the privacy parameters of the algorithm using \texttt{MCMC-DP-Est}, the principal algorithm proposed in~\cite{Yildirim_et_al_2025}. This algorithm uses MCMC sampling to estimate the joint posterior distribution of the privacy parameters.
\end{itemize}

\subsection{Threat Model and Attack Definition}
\label{sec:threat-model}
In our threat model, all involved participants (i.e., all members of the population $\mathcal{P}$ and the server) are assumed to be honest-but-curious. In other words, while they strictly follow the protocol rules, they may attempt to infer private client information from any data available to them. The protocol is resilient to collusion among up to $|\mathcal{P}|-1$ clients, excluding the server. This choice is due to the low likelihood of a server colluding in practical deployments, as they are typically operated by large institutions and are heavily regulated. Nevertheless, in Appendix~\ref{sec:appendix}, we extend this threat model to the setting where the server may collude (with clients).

Our threat model characterizes an adversary as either a participant in the protocol (such as a client or the server) or an external entity that can observe the protocol's actions and outputs. When estimating the privacy of our protocol, we consider the \emph{strongest adversarial view}, denoted as $\mathsf{View}$ in the paper. We compare the information each adversary type can access about the honest client(s)' data to determine the most revealing view, which we then use for privacy estimation. 

In any given round $r$, a group of $|\mathcal{P}|-1$ colluding clients can share their training membership information among themselves (i.e., whether or not they were included in $\mathcal{C}^{(r)}$). Given that the clients also know $|\mathcal{C}^{(r)}|$, this collaboration can ultimately reveal the honest client's identity to the colluding clients. Because of this, standard client MIAs are not appropriate for assessing the protocol's privacy, since the membership information is already inferable within our threat model. To address this, we propose an MIA attack tailored for DP, which estimates the privacy of our protocol by attempting to distinguish the honest client's dataset between two different alternatives.

To formalize the MIA definition tailored for DP and assess the protocol's privacy leakage, we first introduce a \emph{challenge base}.

\begin{definition}[Challenge base] \label{defn: challenge base}
A challenge base refers to a tuple $(\mathcal{D}, \mathcal{Z}_0,\mathcal{Z}_1)$, where $\mathcal{D}$ is a training dataset of $|\mathcal{D}|=K-1$ clients, and $\mathcal{Z}_0$ and $\mathcal{Z}_1$ are the two alternative values of the target client's data.
\end{definition}

Using the definition for a challenge base, we define MIA tailored for DP as follows.

\begin{definition}[MIA tailored for DP] \label{defn: MIAtfDP}
Given a challenge base $(\mathcal{D}, \mathcal{Z}_0,\mathcal{Z}_1)$ and an algorithm $\mathcal{A}$, MIA is a statistical test that aims to determine which of two candidate datasets, $\mathcal{Z}_0$ or $\mathcal{Z}_1$, was used by the target client, by attempting to distinguish between two distributions constructed from the outputs of $\mathcal{A}$ under two different hypotheses:
\[
H_{0}: \theta \sim \mathcal{A}(\mathcal{D} \cup {\mathcal{Z}_{0}}), \quad H_{1}: \theta \sim \mathcal{A}( \mathcal{D} \cup {\mathcal{Z}_{1}}).
\]
\end{definition}
The MIA in Definition \ref{defn: MIAtfDP} has two alternative interpretations. 
\begin{itemize}[leftmargin=*]
\item The first interpretation is related to inferring the target client's data itself. All $|\mathcal{P}| - 1$ clients' data are known (e.g., due to collusion among those clients), and the adversary's intent is to infer the target client's unknown data. The attack in Definition \ref{defn: MIAtfDP} aims to do that when the possible values of the target client's data are reduced to two alternatives, $\mathcal{Z}_{0}$ and $\mathcal{Z}_{1}$. In that way, Definition \ref{defn: MIAtfDP} is different from standard MIA in the literature, where one's membership is inferred. However, this definition is particularly suited to the classical DP setting because it tests the distinguishability of adjacent inputs that differ only in a single client's contribution. 
\item A second interpretation may still be made in terms of actual membership inference. Assume the worst-case scenario, usually adopted in MIA settings, that the data of the whole population, $\{ \mathcal{Z}_{c}: c \in \mP \}$, is known. Assume that the population of clients is partitioned into colluding and non-colluding sets of clients as $\mP = \mP_{\mathsf{C}} \cup \mP_{\mathsf{NC}}$, where there is strictly more than one non-colluding client, i.e., $|\mP_{\mathsf{NC}}| > 1$. Then, a worst-case challenge base in Definition \ref{defn: challenge base} can be constructed where $\mathcal{D} = \{\mathcal{Z}_c : c\in \mathcal{B}\}$, where $\mathcal{B} \subseteq \mP_{\mathsf{C}}$, contains only colluding clients, and $\mathcal{Z}_{0} = \mathcal{Z}_{c_{0}}, \mathcal{Z}_{1} = \mathcal{Z}_{c_{1}}$ where $c_{0}, c_{1} \in \mathcal{P}_{\mathsf{NC}}$ are two non-colluding clients. In other words, the training dataset contains $K-1$ colluding clients and $1$ non-colluding target client. In this case, the attack in Definition \ref{defn: MIAtfDP} can indeed be interpreted as inferring whether the target client is $c_{0}$ or $c_{1}$, which is a genuine membership inference attack. Note that having both $c_{0}$ and $c_{1}$ in $\mP_{\mathsf{NC}}$ is sensible because the membership/non-membership information of all the colluding clients is already known by the attacker.
\end{itemize}

\subsubsection{Training Algorithms}
\label{sec:training-algorithms}

To highlight the benefit of using our proposed method, we compare three federated training algorithms, denoted by $\mathcal{A}_1$, $\mathcal{A}_2$, and $\mathcal{A}_3$, in terms of model utility measured by test loss and estimated privacy:

\paragraph{{${\mathcal{A}_1}$ (DP-only training baseline)}} represents the DP-only FL scenario. At each round, the server generates a global model update by aggregating the perturbed local model updates received from the clients. The server then redistributes the perturbed global update to the next round's participating clients. Training continues from the noisy global update. {Thus, $\mathcal{A}_{1}$ corresponds to a setting in which HE is not used for FL.} 

Alg.~\ref{alg:A1} presents the DP-only training procedure, $\mathcal{A}_1$. We assume each client $c\in\mathcal{C}^{(r)}$ adds a unit Gaussian noise term 
\[
U_c^{(r)} \sim \mathcal{N}\left(0, \sigma^2 I_d\right)
\]
to their local model updates ${\theta}_c^{(r)}$ before sending it to the server. Thus, the conditional variance of the noise on the aggregated global model update $\tilde{\theta}^{(r)}$ given $\{ \tilde\theta_{c}^{(r)}: c \in \mathcal{C}^{(r)} \}$ becomes
\[
 \sigma^{2}\sum_{c \in \mathcal{C}^{(r)} } \alpha_{c}^{2},
\quad
\alpha_c = \frac{n_c}{\sum_{c' \in \mathcal{C}^{(r)}}n_{c'}},
\]
which equates to $\sigma^{2}/K_{s}$ if all $n_{c}:c\in\mathcal{C}^{(r)}$ are equal.

\begin{algorithm}[!h]
\caption{$\mathcal{A}_1$: DP-only training baseline}
\label{alg:A1}

\begin{algorithmic}[1]

\For{$r = 1, \ldots, R$}

    \State Server selects $K_s$ clients
    $\mathcal{C}^{(r)} \subseteq \mathcal{T}$.

    \For{$c \in \mathcal{C}^{(r)}$}

        \State Server sends $\tilde{\theta}^{(r-1)}$ to Client $c$.

        \State Client $c$ obtains
        $\theta_c^{(r)}
        \leftarrow \mathsf{Train}(\tilde{\theta}^{(r-1)}, \mathcal{Z}_c)$.

        \State Client $c$ samples
        $U_c^{(r)} \sim \mathcal{N}(0, \sigma^2 I_d)$.

        \State Client $c$ obtains
        $\tilde{\theta}_c^{(r)}
        = \theta_c^{(r)} + U_c^{(r)}$.

        \State Client $c$ sends $\tilde{\theta}_c^{(r)}$ to Server.

    \EndFor

    \State Server obtains
    $\displaystyle
        \tilde{\theta}^{(r)}
        \leftarrow
        \frac{
            \sum_{c \in \mathcal{C}^{(r)}}
            n_c \tilde{\theta}_c^{(r)}
        }{
            \sum_{c \in \mathcal{C}^{(r)}} n_c
        }.
    $
\EndFor
\end{algorithmic}
\end{algorithm}

\paragraph{{${\mathcal{A}_2}$ (HE training with DP inspection/release)}} represents the HE-based FL scenario with intermittent DP-based inspection, which constitutes our proposed protocol. At each round, the server homomorphically aggregates the encrypted local model updates it receives from the clients. If the global model in that round is to be inspected, the server generates a \emph{separate copy} of the model, adds DP noise, and sends it to the clients. The clients then jointly decrypt the perturbed copy and use it for inspection. The actual training process continues from the non-perturbed encrypted global update. Finally, the server injects DP noise into the final global model and sends it to the clients for decryption. {We stress that the clients continue training with the non-perturbed model without being able to see it, which is enabled in our framework by HE.} We select the variance 
\[
V^{(r)} \sim \mathcal{N}\left(0, \sigma^{2}\sum_{c\in \mathcal{C}^{(r)}} \alpha_c^2 I_d\right)
\] 
to ensure that we match the variance of the fresh Gaussian perturbation in each released global model to establish a common noise baseline across the algorithms. This is equivalent to $V^{(r)} \sim \mathcal{N}(0, \sigma^{2}I_d/K_{s})$ if all $n_{c}:c\in\mathcal{C}^{(r)}$ are equal.

Alg.~\ref{alg:A2} outlines the HE-based training procedure, which applies to both $\mathcal{A}_2$ and $\mathcal{A}_3$, where $\mathcal{I}$ refers to the set of rounds where a separate copy of the global update is made public for inspection/release, which changes depending on the algorithm used. Note that only the clients $c \in \mathcal{T}$ observe the decrypted models, the server and external parties do not.

\begin{algorithm}[!h]
\begin{algorithmic}[1]

\For{$r = 1, \ldots, R$}
    \State Server selects $K_s$ clients $\mathcal{C}^{(r)} \subseteq \mathcal{T}$.

    \For{$c \in \mathcal{C}^{(r)}$}
        \State Server sends $[\theta^{(r-1)}]$ to Client $c$.
        \State Client $c$ obtains $[\theta_c^{(r)}] \gets \mathsf{Train}_{\mathsf{HE}}([\theta^{(r-1)}], \mathcal{Z}_c)$.
        \State Client $c$ sends $[\theta_c^{(r)}]$ to Server.
    \EndFor

    \State Server obtains
    $\displaystyle
    [\theta^{(r)}]
    \gets
    \frac{
        \sum_{c \in \mathcal{C}^{(r)}} n_c [\theta_c^{(r)}]
    }{
        \sum_{c \in \mathcal{C}^{(r)}} n_c
    }$.

    \If{$r \in \mathcal{I}_{\mathcal{A}_{2/3}}$}
        \State Server samples
        $V^{(r)} \sim \mathcal{N}\left(
        0,
        \sigma^2 \sum_{c \in \mathcal{C}^{(r)}} \alpha_c^2 I_d
        \right)$.

        \State Server obtains
        $[\tilde{\theta}^{(r)}]
        \gets
        [\theta^{(r)}] + V^{(r)}$.

        \State Clients $c \in \mathcal{T}$ cooperate to decrypt $\tilde{\theta}^{(r)} \leftarrow [\tilde{\theta}^{(r)}]$.
    \EndIf

\EndFor

\end{algorithmic}
\caption{$\mathcal{A}_{2/3}$: HE-based training with DP inspection/release.}
\label{alg:A2}
\end{algorithm}

\paragraph{{${\mathcal{A}_3}$ (HE training with DP release)}} represents the HE-based FL scenario \emph{without} intermittent inspection. The model is trained under HE throughout all FL rounds, and DP noise is added only to the final global model before downstream release. This algorithm is essentially equivalent to $\mathcal{A}_2$ with intermediate inspection removed.

Fig.~\ref{fig:three_pipelines} presents a simple three-round FL pipeline for all three proposed FL algorithms. Note that, in practice, inspections in $\mathcal{A}_2$ may be performed as often or as rarely as needed, and do not have to occur periodically or on a per-round basis.

\tikzset{
    roundbox/.style={
        draw=OIblue!75!black,
        fill=OIblue!13,
        line width=0.9pt,
        rounded corners=7pt,
        minimum width=3.25cm,
        minimum height=1.50cm,
        align=center,
        inner sep=2pt
    },
    dproundbox/.style={
        draw=OIblue!75!black,
        fill=OIblue!13,
        line width=0.9pt,
        rounded corners=7pt,
        minimum width=3.25cm,
        minimum height=2.0cm,
        align=center,
        inner sep=2pt
    },
    updatebox/.style={
        draw=OIgray,
        fill=black!3,
        line width=0.9pt,
        dash pattern=on 3pt off 3pt,
        rounded corners=5pt,
        minimum width=2.20cm,
        minimum height=1.25cm,
        align=center,
        inner sep=2pt
    },
    inspectbox/.style={
        draw=OIgreen!80!black,
        fill=OIgreen!15,
        line width=0.9pt,
        dash pattern=on 3pt off 3pt,
        rounded corners=5pt,
        minimum width=2.05cm,
        minimum height=1.25cm,
        align=center,
        inner sep=2pt
    },
    finalbox/.style={
        draw=OIvermillion!85!black,
        fill=OIvermillion!18,
        line width=0.9pt,
        dash pattern=on 3pt off 3pt,
        rounded corners=5pt,
        minimum width=2.05cm,
        minimum height=1.25cm,
        align=center,
        inner sep=2pt
    },
    bluearrow/.style={
        -{Stealth[length=3.2mm,width=2.2mm]},
        draw=OIblue,
        line width=1.1pt
    },
    greenarrow/.style={
        -{Stealth[length=3.2mm,width=2.2mm]},
        draw=OIgreen!80!black,
        line width=1.1pt
    },
    redarrow/.style={
        -{Stealth[length=3.2mm,width=2.2mm]},
        draw=OIvermillion,
        line width=1.1pt
    },
    noiselabel/.style={
        font=\scriptsize\sffamily\bfseries,
        align=center,
        inner sep=1pt,
        text=black
    },
    clientmodel/.style={
        draw=OIgray,
        fill=black!2,
        rounded corners=3pt,
        minimum width=1.55cm,
        minimum height=0.58cm,
        align=center,
        inner sep=2pt
    },
    noisebox/.style={
        draw=OIvermillion!75!black,
        fill=OIvermillion!10,
        rounded corners=3pt,
        minimum width=1.35cm,
        minimum height=0.58cm,
        align=center,
        inner sep=2pt
    },
    aggbox/.style={
        draw=OIblue!75!black,
        fill=OIblue!10,
        rounded corners=4pt,
        minimum width=1.55cm,
        minimum height=0.75cm,
        align=center,
        inner sep=2pt,
        font=\scriptsize\sffamily\bfseries
    },
    smallarrow/.style={
        -{Stealth[length=2.3mm,width=1.6mm]},
        draw=OIgray!90!black,
        line width=0.8pt
    }
}
\tikzset{
    roundbox/.style={
        draw=OIblue!75!black,
        fill=OIblue!13,
        line width=0.9pt,
        rounded corners=7pt,
        minimum width=3.25cm,
        minimum height=1.50cm,
        align=center,
        inner sep=2pt
    },
    dproundbox/.style={
        draw=OIblue!75!black,
        fill=OIblue!13,
        line width=0.9pt,
        rounded corners=7pt,
        minimum width=3.25cm,
        minimum height=2.0cm,
        align=center,
        inner sep=2pt
    },
    updatebox/.style={
        draw=OIgray,
        fill=black!3,
        line width=0.9pt,
        dash pattern=on 3pt off 3pt,
        rounded corners=5pt,
        minimum width=2.20cm,
        minimum height=1.25cm,
        align=center,
        inner sep=2pt
    },
    inspectbox/.style={
        draw=OIgreen!80!black,
        fill=OIgreen!15,
        line width=0.9pt,
        dash pattern=on 3pt off 3pt,
        rounded corners=5pt,
        minimum width=2.05cm,
        minimum height=1.25cm,
        align=center,
        inner sep=2pt
    },
    finalbox/.style={
        draw=OIvermillion!85!black,
        fill=OIvermillion!18,
        line width=0.9pt,
        dash pattern=on 3pt off 3pt,
        rounded corners=5pt,
        minimum width=2.05cm,
        minimum height=1.25cm,
        align=center,
        inner sep=2pt
    },
    bluearrow/.style={
        -{Stealth[length=3.2mm,width=2.2mm]},
        draw=OIblue,
        line width=1.1pt
    },
    greenarrow/.style={
        -{Stealth[length=3.2mm,width=2.2mm]},
        draw=OIgreen!80!black,
        line width=1.1pt
    },
    redarrow/.style={
        -{Stealth[length=3.2mm,width=2.2mm]},
        draw=OIvermillion,
        line width=1.1pt
    },
    noiselabel/.style={
        font=\scriptsize\sffamily\bfseries,
        align=center,
        inner sep=1pt,
        text=black
    }
}

\begin{figure*}[!h]
\centering

\begin{minipage}{0.95\linewidth}
\centering

\resizebox{\linewidth}{!}{%
\begin{tikzpicture}[font=\footnotesize\sffamily]

    \node[dproundbox] (a1r1) at (0,0) {
        \textbf{Round 1}\\
        Cleartext Local Training\\
        + Local DP Noise $U_c^{1}$\\
        \& Aggregate
    };

    \node[updatebox] (a1u1) at (4.45,0) {
        Global Model\\
        $\tilde{\theta}^{(1)}$
    };

    \node[dproundbox] (a1r2) at (8.90,0) {
        \textbf{Round 2}\\
        Cleartext Local Training\\
        + Local DP Noise $U_c^{2}$\\
        \& Aggregate
    };

    \node[updatebox] (a1u2) at (13.35,0) {
        Global Model\\
        $\tilde{\theta}^{(2)}$
    };

    \node[dproundbox] (a1r3) at (17.80,0) {
        \textbf{Round 3}\\
        Cleartext Local Training\\
        + Local DP Noise $U_c^{3}$\\
        \& Aggregate
    };

    \node[finalbox] (a1f) at (22.55,0) {
        Final Model\\
        $\tilde{\theta}^{(3)}$
    };

    \draw[bluearrow] (a1r1.east) -- (a1u1.west);
    \draw[bluearrow] (a1u1.east) -- (a1r2.west);
    \draw[bluearrow] (a1r2.east) -- (a1u2.west);
    \draw[bluearrow] (a1u2.east) -- (a1r3.west);
    \draw[redarrow]  (a1r3.east) -- (a1f.west);

\end{tikzpicture}%
}

\small (a) $\mathcal{A}_1$: DP-only training baseline.

\end{minipage}

\vspace{1.2em}

\begin{minipage}{0.95\linewidth}
\centering

\resizebox{\linewidth}{!}{%
\begin{tikzpicture}[font=\footnotesize\sffamily]

    \node[roundbox] (a2r1) at (0,0) {
        \textbf{Round 1}\\
        Encrypted Local\\
        Training \& Agg.
    };

    \node[updatebox] (a2u1) at (4.45,0) {
        Encrypted\\
        Update\\
        $[\theta^{(1)}]$
    };

    \node[roundbox] (a2r2) at (8.90,0) {
        \textbf{Round 2}\\
        Encrypted Local\\
        Training \& Agg.
    };

    \node[updatebox] (a2u2) at (13.35,0) {
        Encrypted\\
        Update\\
        $[\theta^{(2)}]$
    };

    \node[roundbox] (a2r3) at (17.80,0) {
        \textbf{Round 3}\\
        Encrypted Local\\
        Training \& Agg.
    };

    \node[finalbox] (a2f) at (22.55,0) {
        Final Model\\
        $\tilde{\theta}^{(3)}$
    };

    \node[inspectbox] (a2i1) at (4.45,-2.70) {
        Inspection\\
        Model\\
        $\tilde{\theta}^{(1)}$
    };

    \node[inspectbox] (a2i2) at (13.35,-2.70) {
        Inspection\\
        Model\\
        $\tilde{\theta}^{(2)}$
    };

    \draw[bluearrow] (a2r1.east) -- (a2u1.west);
    \draw[bluearrow] (a2u1.east) -- (a2r2.west);
    \draw[bluearrow] (a2r2.east) -- (a2u2.west);
    \draw[bluearrow] (a2u2.east) -- (a2r3.west);
    \draw[redarrow]  (a2r3.east) -- (a2f.west);

    \draw[greenarrow] (a2u1.south) -- (a2i1.north);
    \draw[greenarrow] (a2u2.south) -- (a2i2.north);

    \node[
        noiselabel,
        fill=white,
        anchor=south
    ]
        at ($(a2u1.south)!0.58!(a2i1.north)+(0,0.02)$)
        {+ Noise $V_1$ \& Decrypt};

    \node[
        noiselabel,
        fill=white,
        anchor=south
    ]
        at ($(a2u2.south)!0.58!(a2i2.north)+(0,0.02)$)
        {+ Noise $V_2$ \& Decrypt};

    \node[
        noiselabel,
        anchor=south,
        text width=2.10cm
    ]
        at ($(a2r3.east)!0.50!(a2f.west)+(0,0.05)$)
        {+ Noise $V_3$\\[-1pt]\& Decrypt};

\end{tikzpicture}%
}

\small (b) $\mathcal{A}_2$: HE-based training with DP inspection.

\end{minipage}

\vspace{1.2em}

\begin{minipage}{0.95\linewidth}
\centering

\resizebox{\linewidth}{!}{%
\begin{tikzpicture}[font=\footnotesize\sffamily]

    \node[roundbox] (a3r1) at (0,0) {
        \textbf{Round 1}\\
        Encrypted Local\\
        Training \& Agg.
    };

    \node[updatebox] (a3u1) at (4.45,0) {
        Encrypted\\
        Update\\
        $[\theta^{(1)}]$
    };

    \node[roundbox] (a3r2) at (8.90,0) {
        \textbf{Round 2}\\
        Encrypted Local\\
        Training \& Agg.
    };

    \node[updatebox] (a3u2) at (13.35,0) {
        Encrypted\\
        Update\\
        $[\theta^{(2)}]$
    };

    \node[roundbox] (a3r3) at (17.80,0) {
        \textbf{Round 3}\\
        Encrypted Local\\
        Training \& Agg.
    };

    \node[finalbox] (a3f) at (22.55,0) {
        Final Model\\
        $\tilde{\theta}^{(3)}$
    };

    \draw[bluearrow] (a3r1.east) -- (a3u1.west);
    \draw[bluearrow] (a3u1.east) -- (a3r2.west);
    \draw[bluearrow] (a3r2.east) -- (a3u2.west);
    \draw[bluearrow] (a3u2.east) -- (a3r3.west);
    \draw[redarrow]  (a3r3.east) -- (a3f.west);

    \node[
        noiselabel,
        anchor=south,
        text width=2.10cm
    ]
        at ($(a3r3.east)!0.50!(a3f.west)+(0,0.05)$)
        {+ Noise $V_3$\\[-1pt]\& Decrypt};

\end{tikzpicture}
}

\small (c) $\mathcal{A}_3$: HE-based training with final DP release only.

\end{minipage}

\vspace{0.4em}

\caption{Overview of the three evaluated FL algorithms.}
\label{fig:three_pipelines}
\end{figure*}
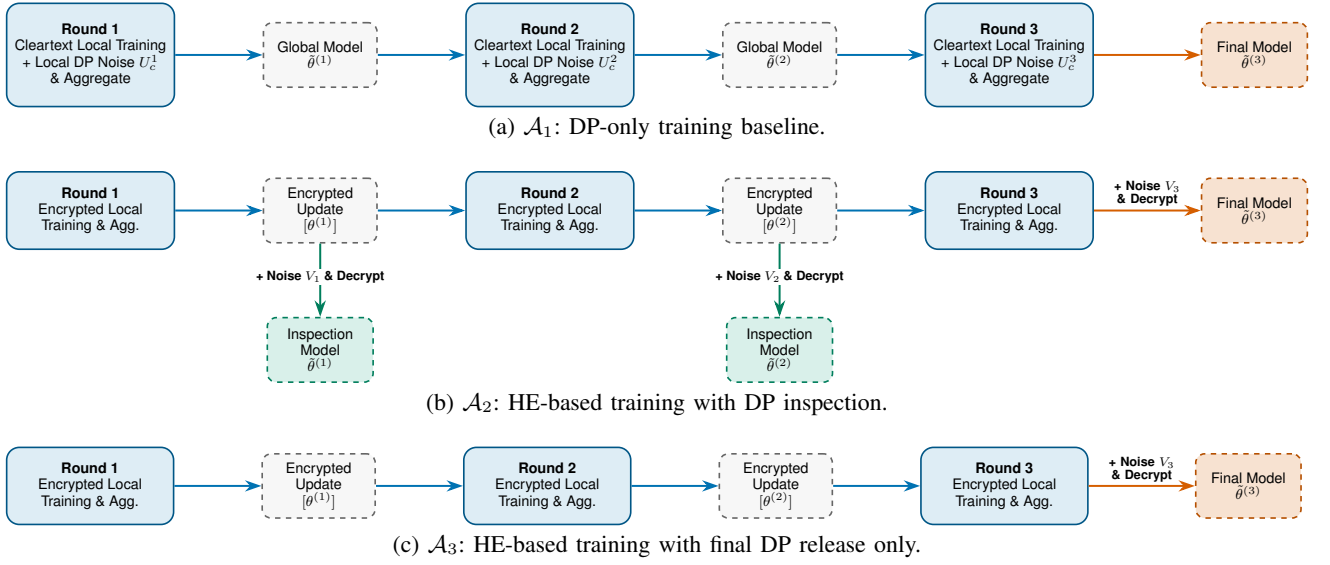

\subsubsection{Adversary's View}
\label{sec:adversarys-view}

In this section, we determine the maximum information an adversary can extract about the honest client $c_h$'s data from a single federated learning run under our threat model and using our training algorithms. This information is used directly for privacy estimation, since it captures the worst-case information an adversary can obtain from a single run under our defined threat model.

\paragraph{$\mathcal{A}_1$ (DP-only training baseline)}
By sharing their membership information with each other, colluding clients (i.e., all clients in $\mathcal{P}$ other than the honest client $c_h$) can determine precisely, among the $R$ rounds, the set of rounds in which $c_h$ participated:
$
J_{c_h}^{(R)} = \{ r \in \{1, \ldots, R\}: c_h \in \mathcal{C}^{(r)} \}.
$
Furthermore, since each colluding client knows its own local updates and noise terms, they can isolate $c_h$'s noisy local updates. Hence, the strongest adversarial view, i.e., the information available and inferred by the colluding clients, can be represented as in Eq.~\ref{eq: A1 adv view}.
\begin{equation}
 \mathsf{View}_{\mathcal{A}_1}^{(R)} = \{ \tilde{\theta}_{c_h}^{(r)}: r \in J_{c_h}^{(R)}, c_h \in \mathcal{C}^{(r)}  \}
\label{eq: A1 adv view}
\end{equation}

\paragraph{$\mathcal{A}_{2/3}$ (HE-based training)}
Unlike $\mathcal{A}_1$, colluding clients cannot directly extract $c_h$'s noisy updates, $\tilde\theta^{(r)}_{c_h}$. In $\mathcal{A}_{2/3}$, the entire training is realized homomorphically. As a result, clients cannot directly access the plaintext values of their local updates, since these are also computed homomorphically using the \emph{encrypted and unperturbed} global model update received from the previous round. Furthermore, if any client wishes to decrypt their local updates, all clients involved in training (i.e., $\mathcal{T}$, which includes $c_h$) must agree to perform the decryption, which prevents the release of local models. Because no one can directly access their local updates or the noise term added by the non-colluding server, the most information an adversary can access regarding $\mathcal{Z}_{c_h}$ remains the perturbed global model update \emph{at inspection/release rounds}. Hence, the strongest adversarial view can
be represented as in Eq.~\ref{eq: A2 adv view}, where $\mathcal{I}^{(R)}_{\mathcal{A}_2} \subseteq \{1, \ldots, R\}$ and $\mathcal{I}^{(R)}_{\mathcal{A}_3} = \{R\}$ denote the inspection/release rounds for $\mathcal{A}_2$ and $\mathcal{A}_3$ respectively.
\begin{equation}
    \mathsf{View}_{\mathcal{A}_{2/3}}^{(R)} = \{ \tilde{\theta}^{(r)}: r \in \mathcal{I}^{(R)}_{\mathcal{A}_{2/3}} \}
\label{eq: A2 adv view}
\end{equation}

We observe that the adversary's view per round is noticeably noisier in $\mathcal{A}_{2/3}$ when we target the same scale of individual client contributions in the revealed models.
Assume for simplicity that $n_{c}$s are the same. Since 
\[
\tilde{\theta}^{(r)} = \frac{1}{K_{s}} \sum_{c \in \mathcal{C}_{r}} {\theta}_{c}^{(r)} + V^{(r)} = \left[ \sum_{c \in \mathcal{C}_{r}} {\theta}_{c}^{(r)} + K_{s} V^{(r)} \right]/K_{s},
\]
the appropriately scaled version of each individual's contribution (including $c_h$'s) can be observed with noise that has a variance of $K_{s}^{2} \sigma^{2}/K_{s} = K_{s} \sigma^{2}$. This analysis indicates that noise introduced by a non-colluding server offers stronger privacy protection.

\subsubsection{MIA tailored for DP}
\label{sec:miatfdp}

To apply the MIA-based privacy estimation method in our FL setting, we need to define the corresponding test statistic. For each challenge base $(\mathcal{D}, \mathcal{Z}_{0}, \mathcal{Z}_{1})$, we run the training algorithm $\mathcal{A}$, under two hypotheses, $H_0$ and $H_1$.
Under $H_0$, the data of the target individual, who corresponds to the honest client $c_h$ in our protocol's threat model, is selected as $\mathcal{Z}_{0}$ (i.e., the training set is $\mathcal{D}\cup\mathcal{Z}_{0}$); whereas under $H_1$, the target individual's data is $\mathcal{Z}_{1}$ (i.e., the training set is $\mathcal{D}\cup\mathcal{Z}_{1}$). To estimate the privacy for each algorithm $\mathcal{A}$, we utilize the $\mathsf{View}$ defined in Section~\ref{sec:adversarys-view}, providing the attacker with the most informative model sequence revealed about the target client.

We train $N$ shadow models each under $H_{0}$ and $H_{1}$ for $R$ rounds. For each shadow execution $i \in \{1, \ldots, N\}$ under $H_b$ where $b \in \{0, 1\}$, we compute the test statistics
\[
L_{b,\mathcal{A}_1, i}^{(R)} 
= 
\frac{1}{\left|J_{c_b,i }^{(R)}\right|}
\sum_{r \in J_{c_b, i}^{(R)}} 
\ell\!\left(\tilde{\theta}_{b, c_b,i }^{(r)}, \mathcal{Z}_{0}\right),
\]
\[
L_{b,\mathcal{A}_{2/3},i}^{(R)} 
= 
\frac{1}{\left|\mathcal{I}^{(R)}_{\mathcal{A}_{2/3}}\right|}
\sum_{r \in \mathcal{I}^{(R)}_{\mathcal{A}_{2/3}}} 
\ell\!\left(\tilde{\theta}_{b,i}^{(r)}, \mathcal{Z}_{0}\right).
\]
\noindent The attacker learns the distributions for the hypotheses $H_{0}$ and $H_{1}$ from the test statistics $\{ L_{0,\mathcal{A},i}^{(R)} \}$ and $\{ L_{1,\mathcal{A},i}^{(R)} \}$, respectively.

Given the adversary's view up to round $R$ of the \emph{actual} learning process (i.e., not the shadow processes), the attacker calculates the observed statistic
\[
L_{\mathcal{A}_1}^{(R)}
=
\frac{1}{\left|J_{c_h}^{(R)}\right|}
\sum_{r \in J_{c_h}^{(R)}}
\ell\!\left(\tilde{\theta}_{c_h}^{(r)}, \mathcal{Z}_{0}\right),
\]
\[
L_{\mathcal{A}_{2/3}}^{(R)} 
= 
\frac{1}{\left|\mathcal{I}^{(R)}_{\mathcal{A}_{2/3}}\right|}
\sum_{r \in \mathcal{I}^{(R)}_{\mathcal{A}_{2/3}}} 
\ell\!\left(\tilde{\theta}^{(r)}, \mathcal{Z}_{0}\right).
\]
The attacker compares the observed statistic with the distributions derived for $H_0$ and $H_1$ to determine whether the actual training data of the target client corresponds to $\mathcal{Z}_{0}$ or $\mathcal{Z}_{1}$. For more information about how we learn the distributions for $H_0$ and $H_1$ and apply a decision, refer to the \texttt{LearnAndDecide} algorithm in~\cite{Yildirim_et_al_2025}. 

The notation $\ell(\theta, \mathcal{Z})$ represents the unsmoothed loss of the model $\theta$ on the dataset $\mathcal{Z}$. When calculating the test statistic, we fix the second parameter at $\mathcal{Z}_{0}$. This ensures the loss is evaluated with respect to the same dataset for both hypotheses, enabling a comparable analysis of the underlying distributions.

\section{Experimental Results}
\label{sec:experimental-results}
\subsection{Dataset}
\label{sec:dataset-setup}
In our experiments, we use the Federated Extended MNIST (FEMNIST) dataset from LEAF~\cite{caldas2018leaf}. LEAF is a benchmarking framework specifically designed for FL settings, providing open-source datasets constructed to reflect the characteristics of federated environments realistically. Similarly, in the FEMNIST dataset, the data is naturally partitioned across clients in a heterogeneous fashion, and the number of samples assigned to each client differs considerably. This setup captures the non-i.i.d. and unbalanced data distribution typically observed in practical FL scenarios.

We use a random subset of the FEMNIST dataset as the population~$\mathcal{P}$. This subset consists of $188$ distinct clients (handwriters), each contributing handwriting samples across a subset of $62$ distinct characters. Each individual's data is split into training and test subsets with a training ratio of $0.8$.

\subsection{FL Setup}
In the established privacy scenario and proposed FL setting, we select a training cohort $\mathcal{T}$ of $100$ participants from the population $\mathcal{P}$ for model training. In each round $r$ of FL, the server randomly selects a subset of the clients $\mathcal{C}^{(r)} \subseteq \mathcal{T}$ of size  $K_s=5$  and only the local updates of the selected clients are used to generate the global update of that round. The number of global FL rounds, $R$, is set to $500$, and the number of local epochs is set to $1$. During local training, the batch size is set to $16$. We use Stochastic Gradient Descent~(SGD) as the optimizer with a momentum of $0.9$. We set the learning rate, $\mu$, to $0.01$ and apply label smoothing with a coefficient of $0.01$ to stabilize training. The same smoothed loss is also utilized when reporting the test loss for utility evaluation in Section~\ref{sec:experimental-results}.

The model we use is a lightweight convolutional neural network~(CNN). The architecture of the model is summarized in Table~\ref{tab:cnn_architecture}.

\begin{table}[h]
\centering
\caption{CNN architecture used for FEMNIST classification.}
\label{tab:cnn_architecture}
\begin{tabular}{lll}
\toprule
Layer & Configuration & Output shape \\
\midrule
Input & Grayscale FEMNIST image & $28 \times 28 \times 1$ \\
Conv2D + ReLU & 32 filters, $3 \times 3$ kernel & $26 \times 26 \times 32$ \\
AvgPool2D & $2 \times 2$ pool size & $13 \times 13 \times 32$ \\
Conv2D + ReLU & 64 filters, $3 \times 3$ kernel & $11 \times 11 \times 64$ \\
AvgPool2D & $2 \times 2$ pool size & $5 \times 5 \times 64$ \\
Flatten & -- & $1600$ \\
Dense + ReLU & 512 units & $512$ \\
Dense + Softmax & 62 units & $62$ \\
\bottomrule
\end{tabular}
\end{table}

\subsection{Attack Evaluation}
\label{sec:attack-evaluation}
\subsubsection{Challenge Construction}
\label{sec:challenge-construction}

In our experiments, we generate a total of $n=20$ challenge bases, and each challenge base is constructed {such that $\mathcal{D}_{i} \cup \mathcal{Z}_{i, 0}$ and $\mathcal{D}_{i} \cup \mathcal{Z}_{i, 1}$ includes $100$ individuals' data, where $i \in \{1, \ldots, n\}$.} For each challenge base, we train $N$ shadow models under each hypothesis. To measure the attack performance, we use the leave-one-out technique (see \texttt{MeasureMIAFast} in~\cite{Yildirim_et_al_2025}).

In our experiments, the unit noise generated by clients (and added to the local models) in $\mathcal{A}_1$ is zero-mean Gaussian noise with a fixed standard deviation of $\sigma=0.02$, unless specified otherwise. For the algorithms $\mathcal{A}_{2/3}$, the standard deviation of the noise the server adds to the global models is determined based on the derivations discussed in Section~\ref{sec:adversarys-view}, and it varies depending on the number of data points held by each client in the subset $c \in \mathcal{C}^{(r)}$. Although the variance of the noise introduced to the global model update at each round remains consistent across all algorithms ($ \mathcal{A}_{1/2/3}$), it can differ from one round to the next. We present the results starting from round $100$, as earlier rounds do not provide reliable privacy estimates due to our attack not being the strongest or most ideal.

\subsection{Results}
\subsubsection{Privacy and Effects on Utility}
\label{sec:priv-eff-on-acc}

One of our main claims is that $\mathcal{A}_3$ can achieve higher estimated privacy than $\mathcal{A}_1$ while maintaining the same or improved levels of utility. We utilize our MCMC-based privacy estimation procedure to calculate the posterior mean of $\epsilon$, focusing on the most informative models for the target client that an adversary can extract from the protocol outputs available in its view. 

For ${\mathcal A}_3$, we report the estimated privacy at each round by treating the model at that round as the final released model. In other words, each point on the {\tt x}-axis corresponds to the privacy estimate if training were terminated and the DP-perturbed model were released at that round. 

For utility evaluations, we select a single challenge base under $H_0$ and report the average test loss across its $N=100$ associated shadow models evaluated on the concatenated local test sets of the clients in $H_0$, since the full set of challenge bases and hypotheses is required only for privacy estimation.

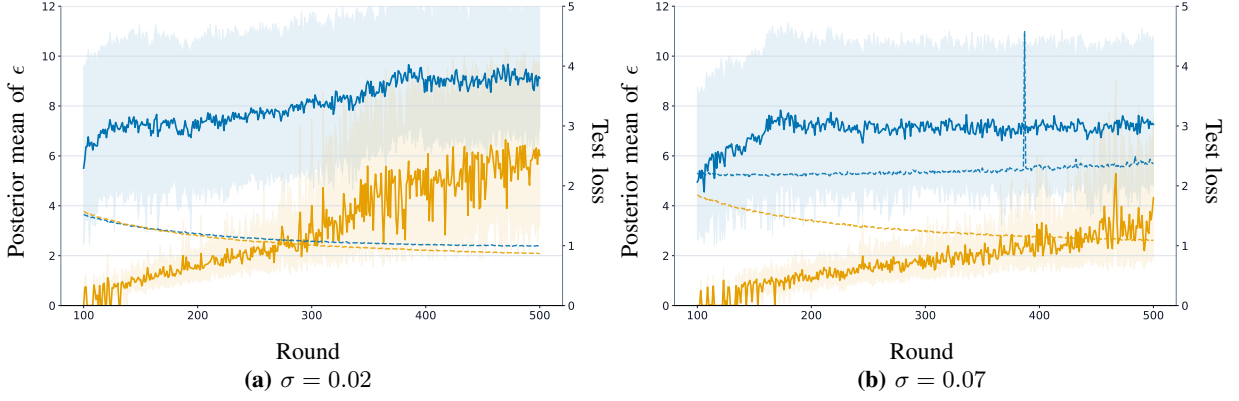
\begin{figure*}[!h]
    \centering
    \begin{tikzpicture}
        \node[inner sep=0] (img1) {
            \includegraphics[
                width=0.40\textwidth
            ]{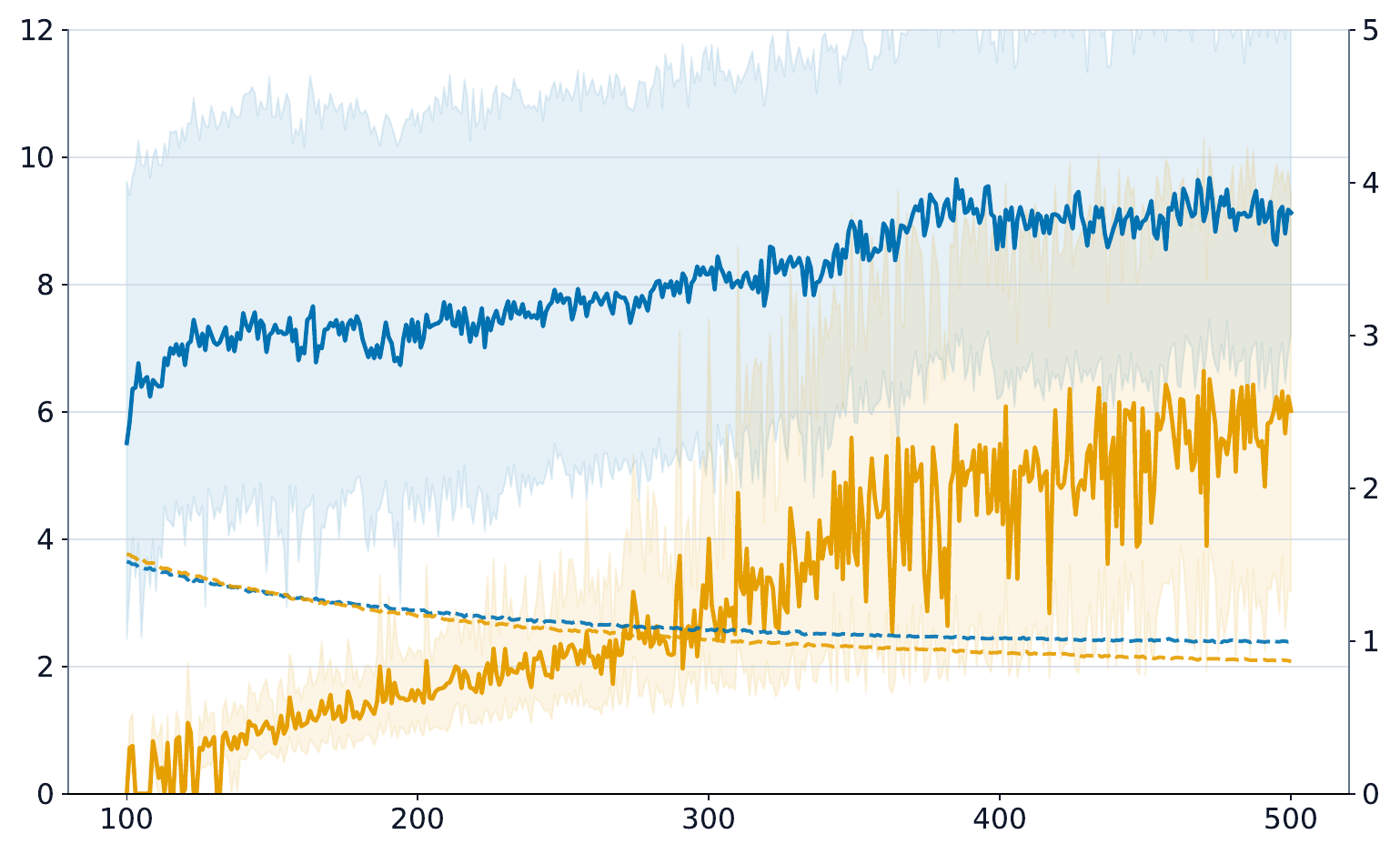}
        };

        \node[rotate=90, anchor=south, font=\small]
            at (img1.west)
            {Posterior mean of $\epsilon$};

        \node[rotate=270, anchor=south, font=\small]
            at (img1.east)
            {Test loss};

        \node[anchor=north, font=\small]
            at ([yshift=-0.05cm]img1.south)
            {Round};

        \node[anchor=north, font=\small]
            at ([yshift=-0.42cm]img1.south)
            {\textbf{(a)} $\sigma = 0.02$};

        \node[inner sep=0, anchor=west] (img2)
            at ([xshift=0.85cm]img1.east) {
            \includegraphics[
                width=0.40\textwidth
            ]{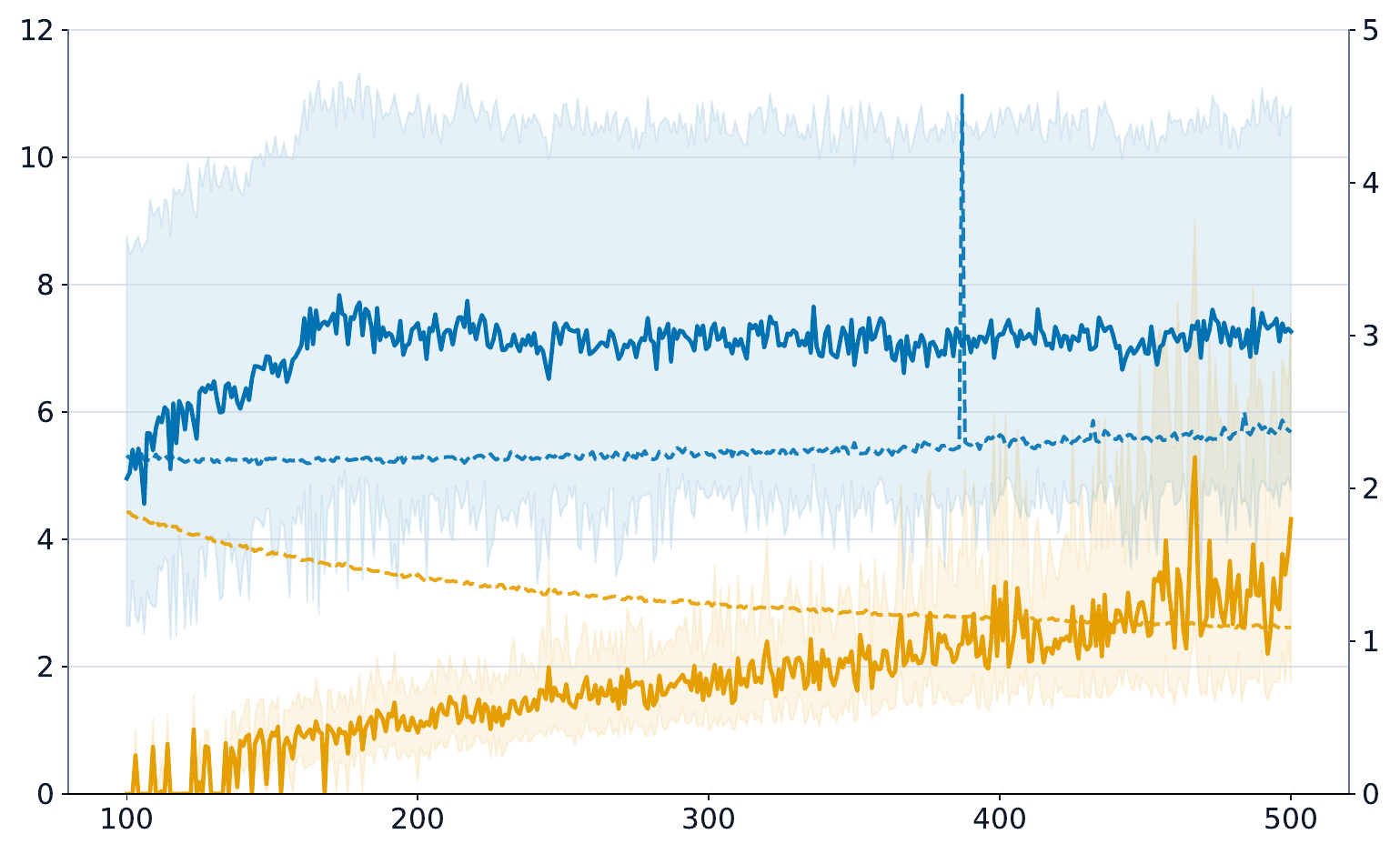}
        };

        \node[rotate=90, anchor=south, font=\small]
            at (img2.west)
            {Posterior mean of $\epsilon$};

        \node[rotate=270, anchor=south, font=\small]
            at (img2.east)
            {Test loss};

        \node[anchor=north, font=\small]
            at ([yshift=-0.05cm]img2.south)
            {Round};

        \node[anchor=north, font=\small]
            at ([yshift=-0.42cm]img2.south)
            {\textbf{(b)} $\sigma = 0.07$};

    \end{tikzpicture}

    \caption{Privacy--utility comparison between $\mathcal{A}_1$
    and $\mathcal{A}_3$ for (a) $\sigma=0.02$ and (b) $\sigma=0.07$.
    Blue and orange denote \textcolor{OIblue}{$\mathcal{A}_1$} and
    \textcolor{OIorange}{$\mathcal{A}_3$}, respectively; solid and dashed
    lines show posterior mean $\epsilon$ and mean loss, with 95\% credible
    intervals shaded.}
    \label{fig:loss_vs_eps}
\end{figure*}

As shown in Fig.~\ref{fig:loss_vs_eps}(a), at round $500$, for comparable levels of utility, corresponding to test losses of $1.00$ for $\mathcal{A}_1$ and $0.86$ for $\mathcal{A}_3$, the estimated privacy loss of $\mathcal{A}_1$ is noticeably higher than that of $\mathcal{A}_3$, with posterior mean estimates of $\epsilon$ of $9.13$ and $6.02$, respectively. This trend persists throughout the remainder of the training process, supporting our argument that training under HE and applying DP only at model release yields lower estimated privacy loss at comparable utility than perturbing the global update each round and continuing training on the perturbed model. Moreover, Fig.~\ref{fig:loss_vs_eps}(b) demonstrates that even at higher noise levels (when required to achieve stronger privacy), at round $500$, $\mathcal{A}_3$ can attain better utility than $\mathcal{A}_1$, with losses of $1.09$ and $2.37$, respectively, while still preserving a large improvement in estimated privacy, with posterior mean estimates of $\epsilon$ of $4.32$ and $7.26$, respectively. This trend again persists throughout the remainder of the training process.

\begin{figure*}[!h]
\centering
\begin{tikzpicture}
\node[inner sep=0] (grid) {
\begin{minipage}{\linewidth}
\centering

    \hspace{-1cm}
    \begin{minipage}{0.29\linewidth}
        \centering
        \includegraphics[width=\linewidth]{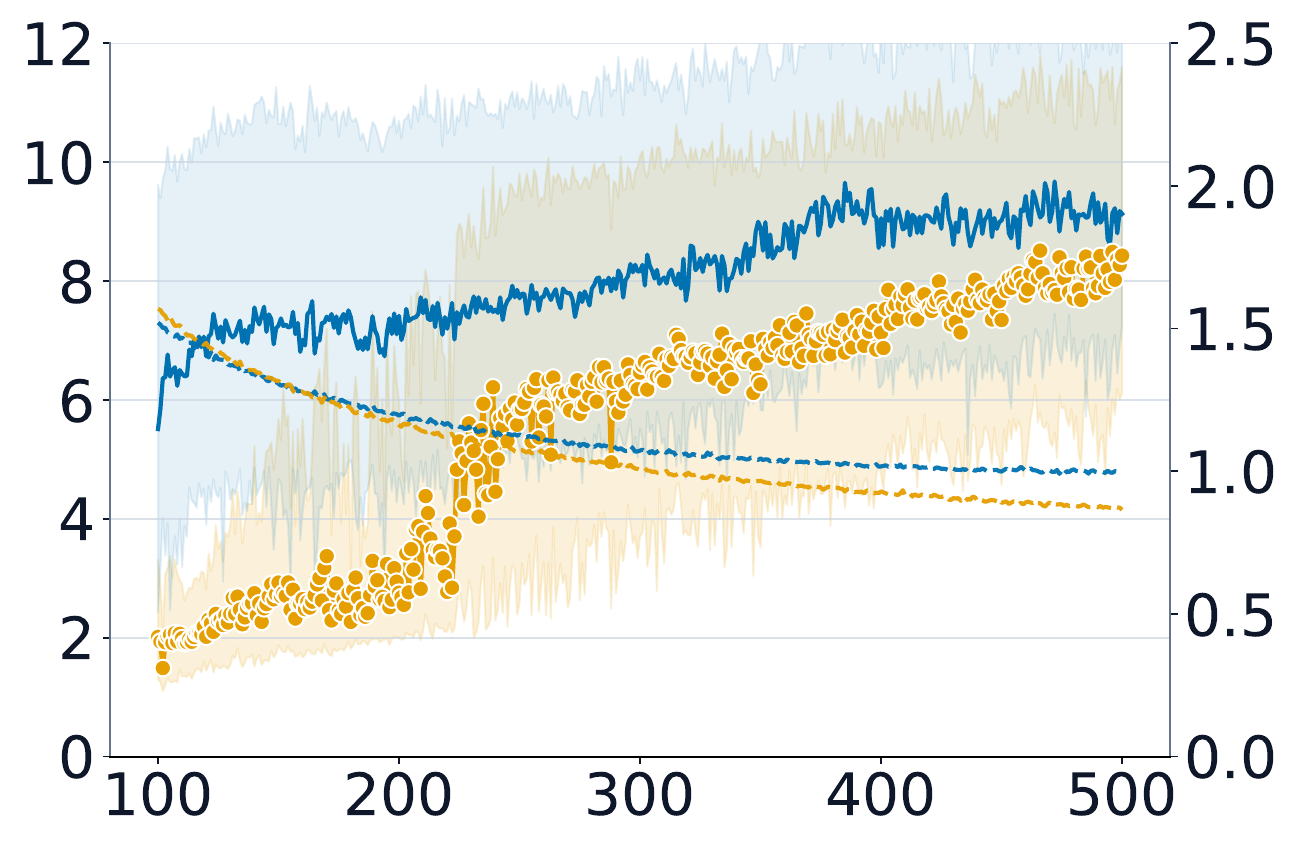}\\[-0.4em]
        {\footnotesize $P=1$}
    \end{minipage}\hspace{0.1cm}
    \begin{minipage}{0.29\linewidth}
        \centering
        \includegraphics[width=\linewidth]{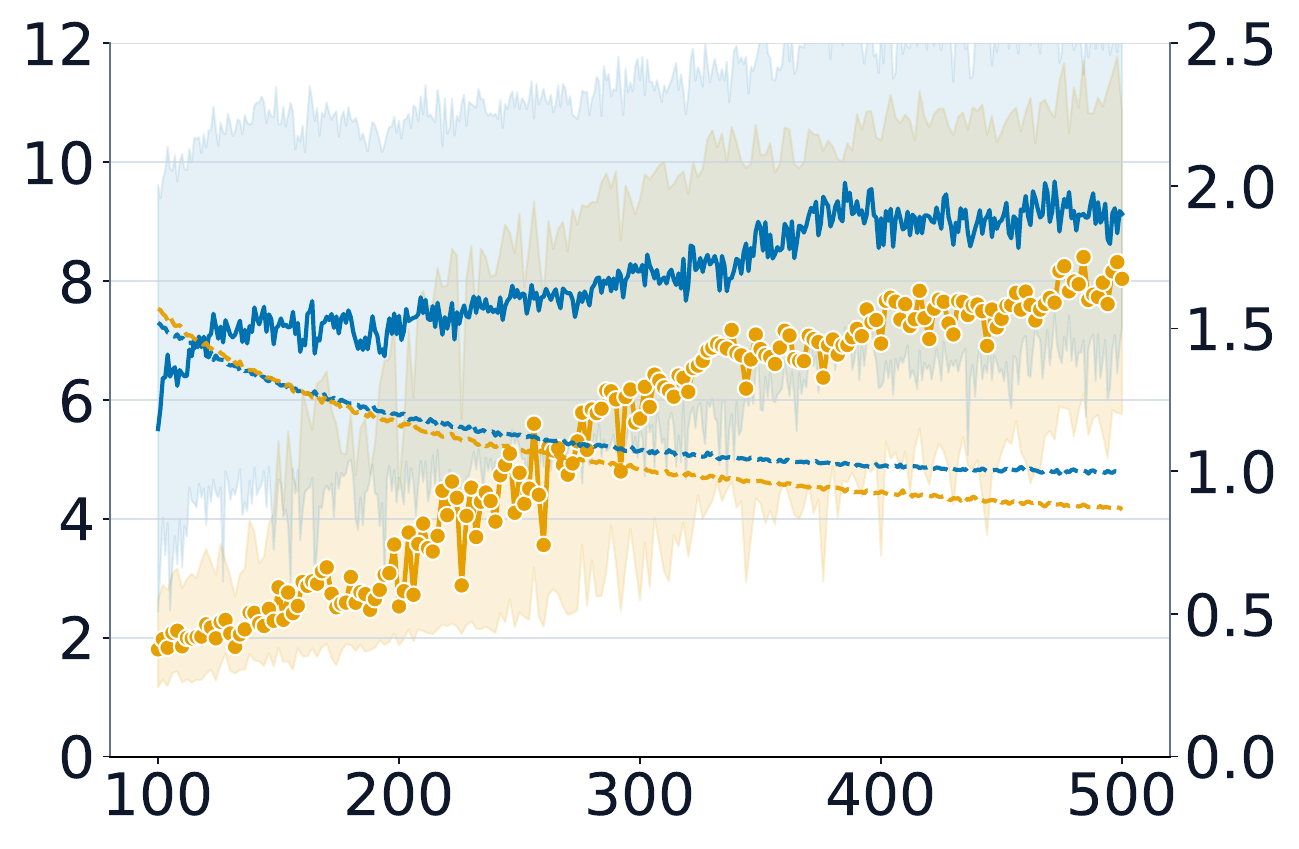}\\[-0.4em]
        {\footnotesize $P=2$}
    \end{minipage}\hspace{0.1cm}
    \begin{minipage}{0.29\linewidth}
        \centering
        \includegraphics[width=\linewidth]{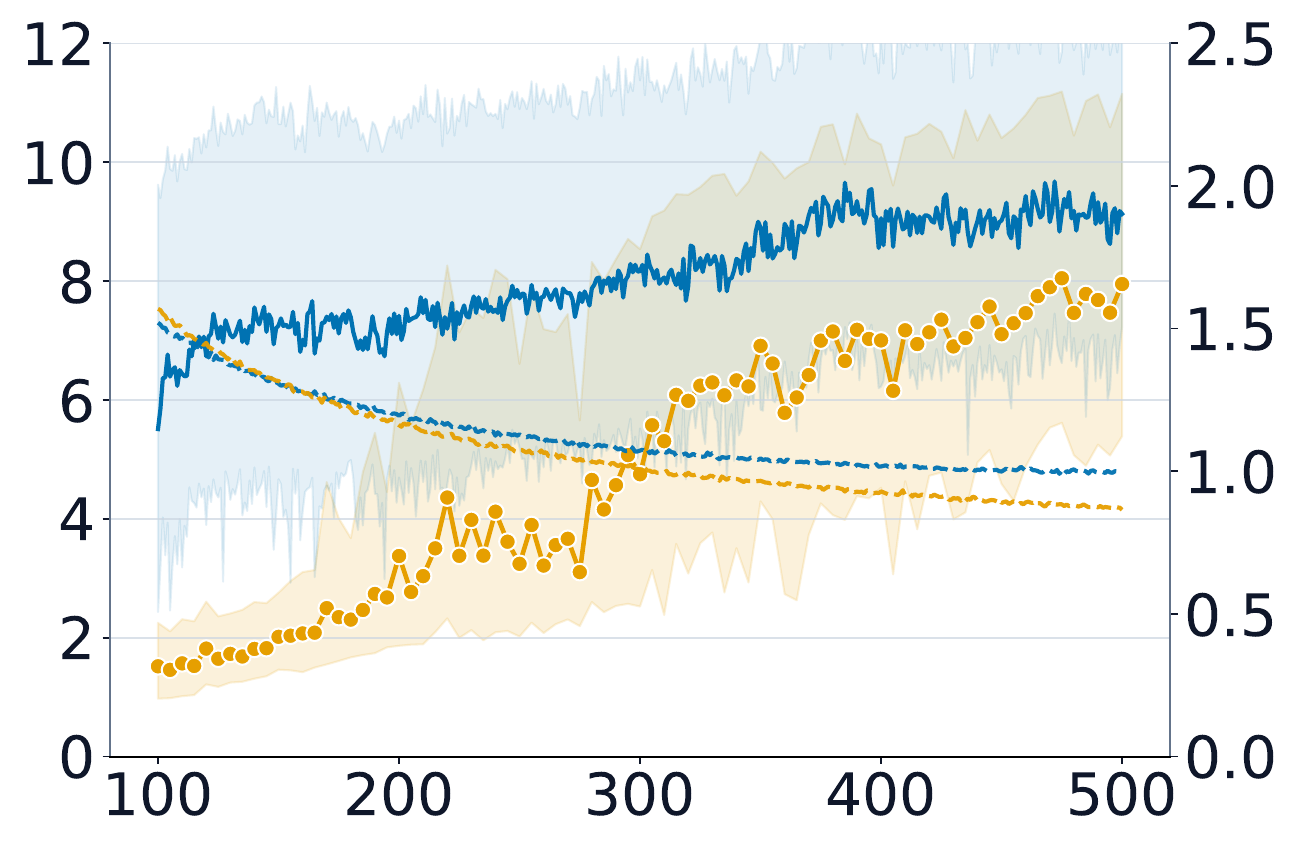}\\[-0.4em]
        {\footnotesize $P=5$}
    \end{minipage}

    \vspace{0.2em}

    \hspace{-1cm}
    \begin{minipage}{0.29\linewidth}
        \centering
        \includegraphics[width=\linewidth]{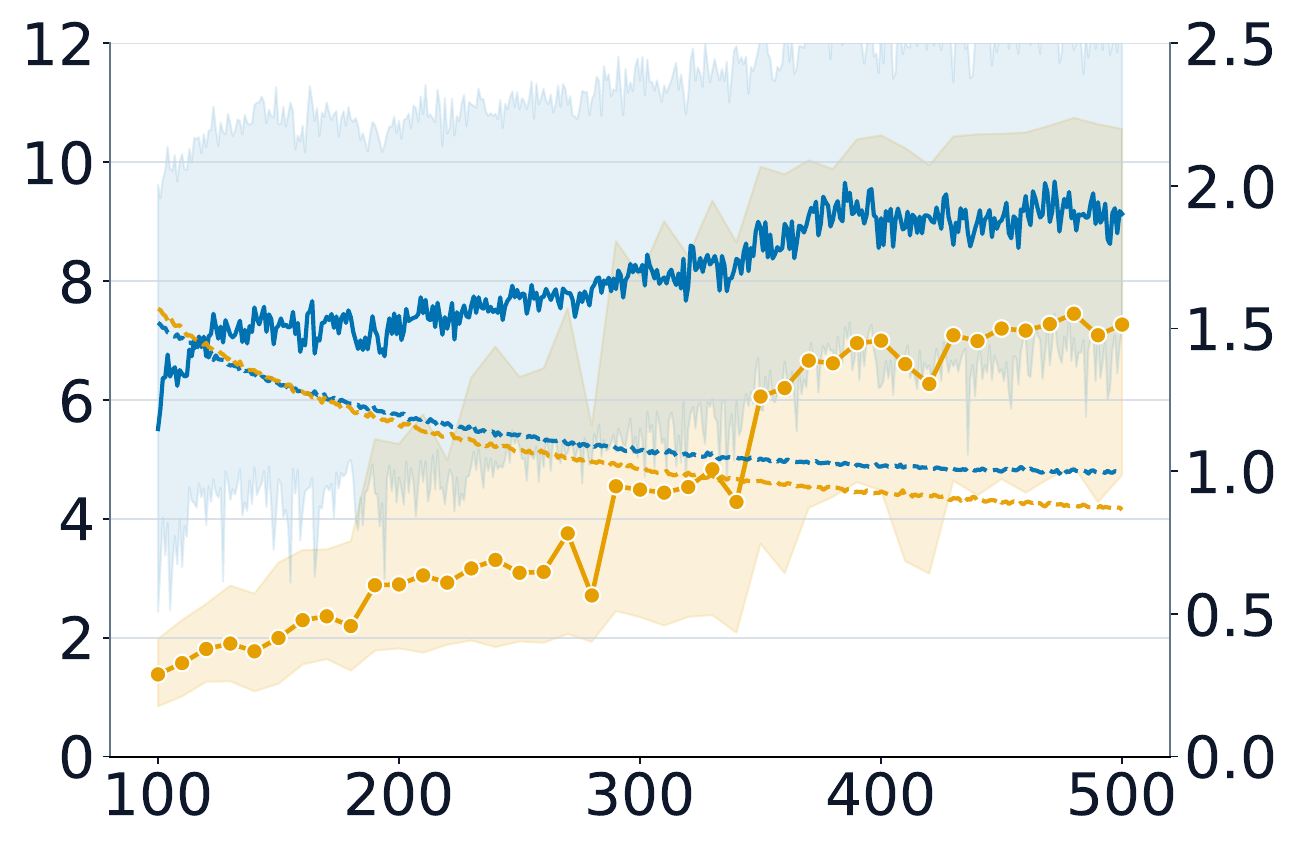}\\[-0.4em]
        {\footnotesize $P=10$}
    \end{minipage}\hspace{0.1cm}
    \begin{minipage}{0.29\linewidth}
        \centering
        \includegraphics[width=\linewidth]{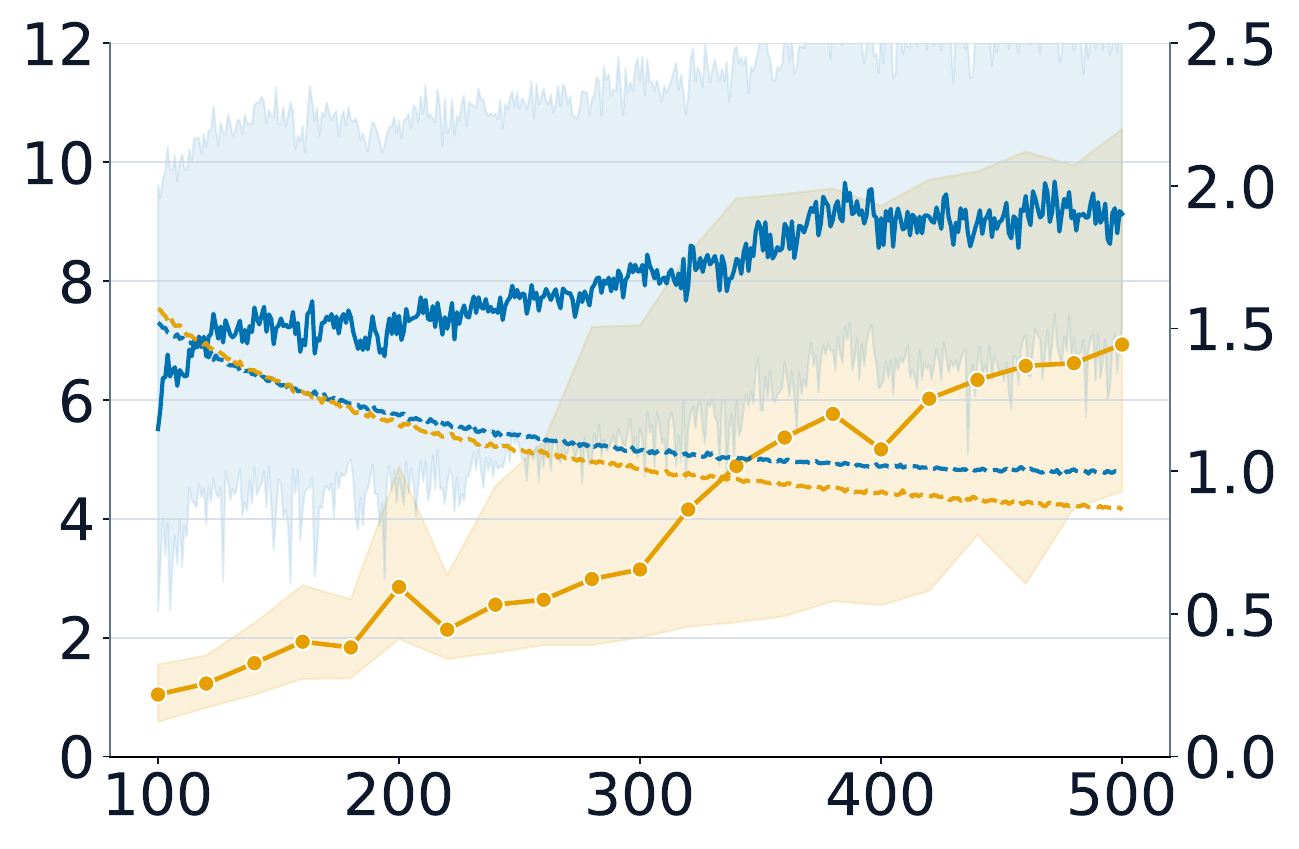}\\[-0.4em]
        {\footnotesize $P=20$}
    \end{minipage}\hspace{0.1cm}
    \begin{minipage}{0.29\linewidth}
        \centering
        \includegraphics[width=\linewidth]{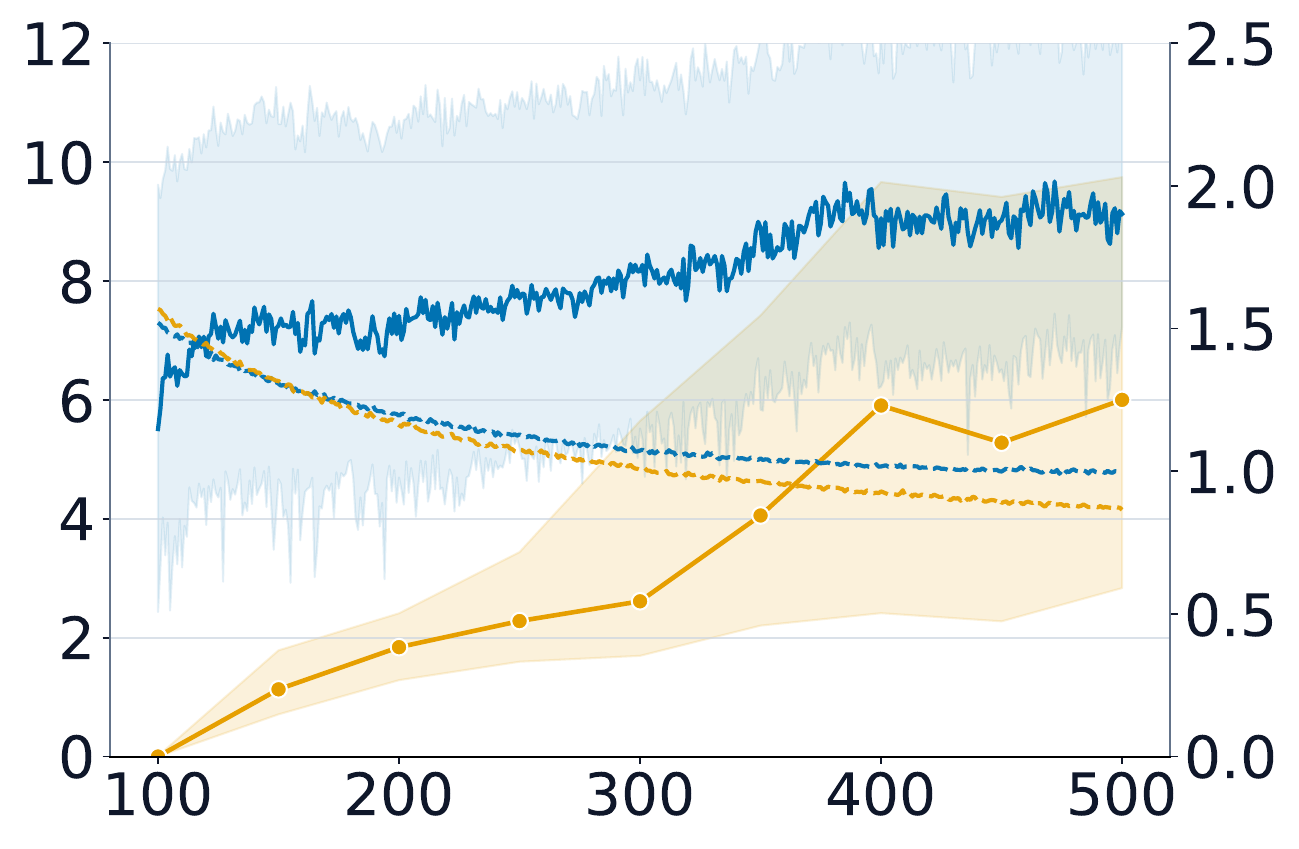}\\[-0.4em]
        {\footnotesize $P=50$}
    \end{minipage}

    \vspace{0.2em}

    \hspace{-1cm}
    \begin{minipage}{0.29\linewidth}
        \centering
        \includegraphics[width=\linewidth]{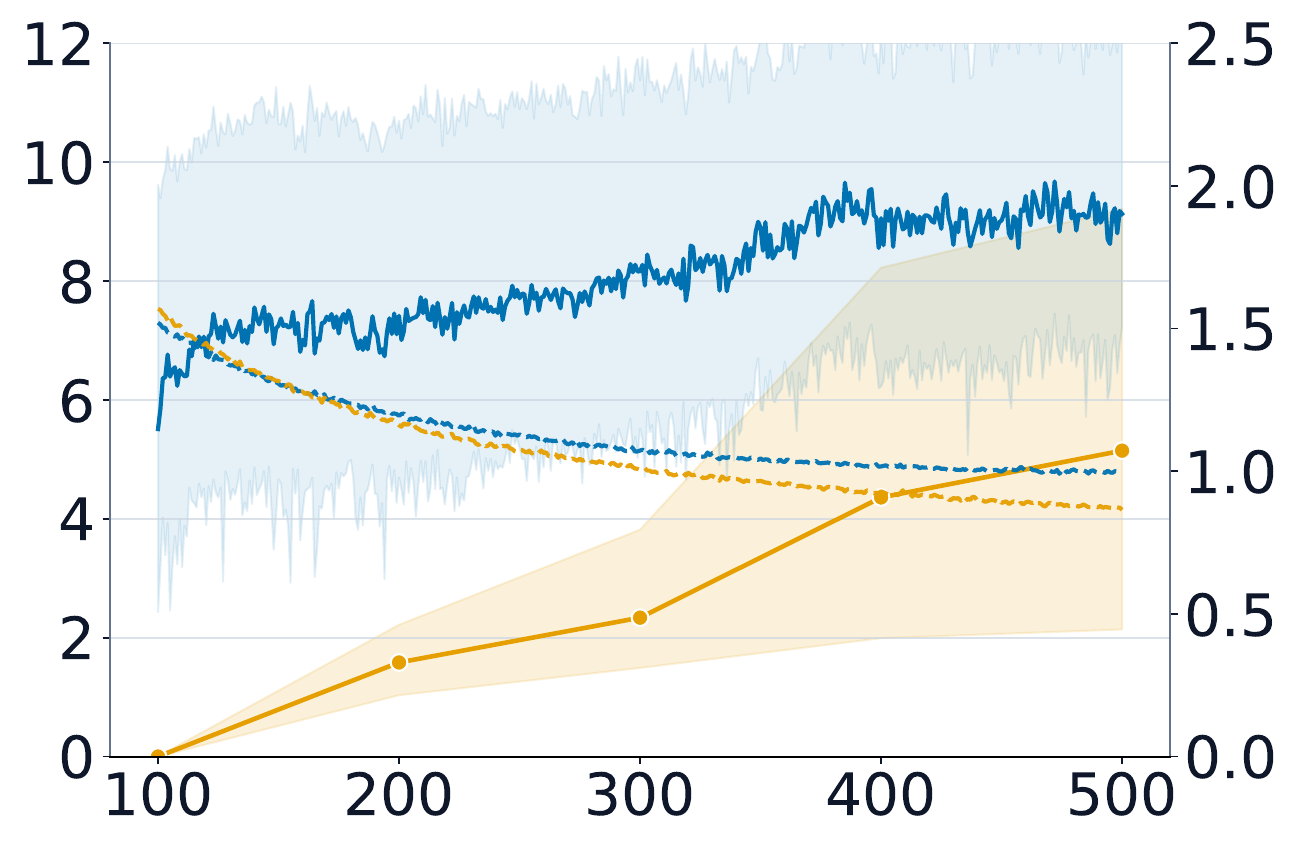}\\[-0.4em]
        {\footnotesize $P=100$}
    \end{minipage}\hspace{0.1cm}
    \begin{minipage}{0.29\linewidth}
        \centering
        \includegraphics[width=\linewidth]{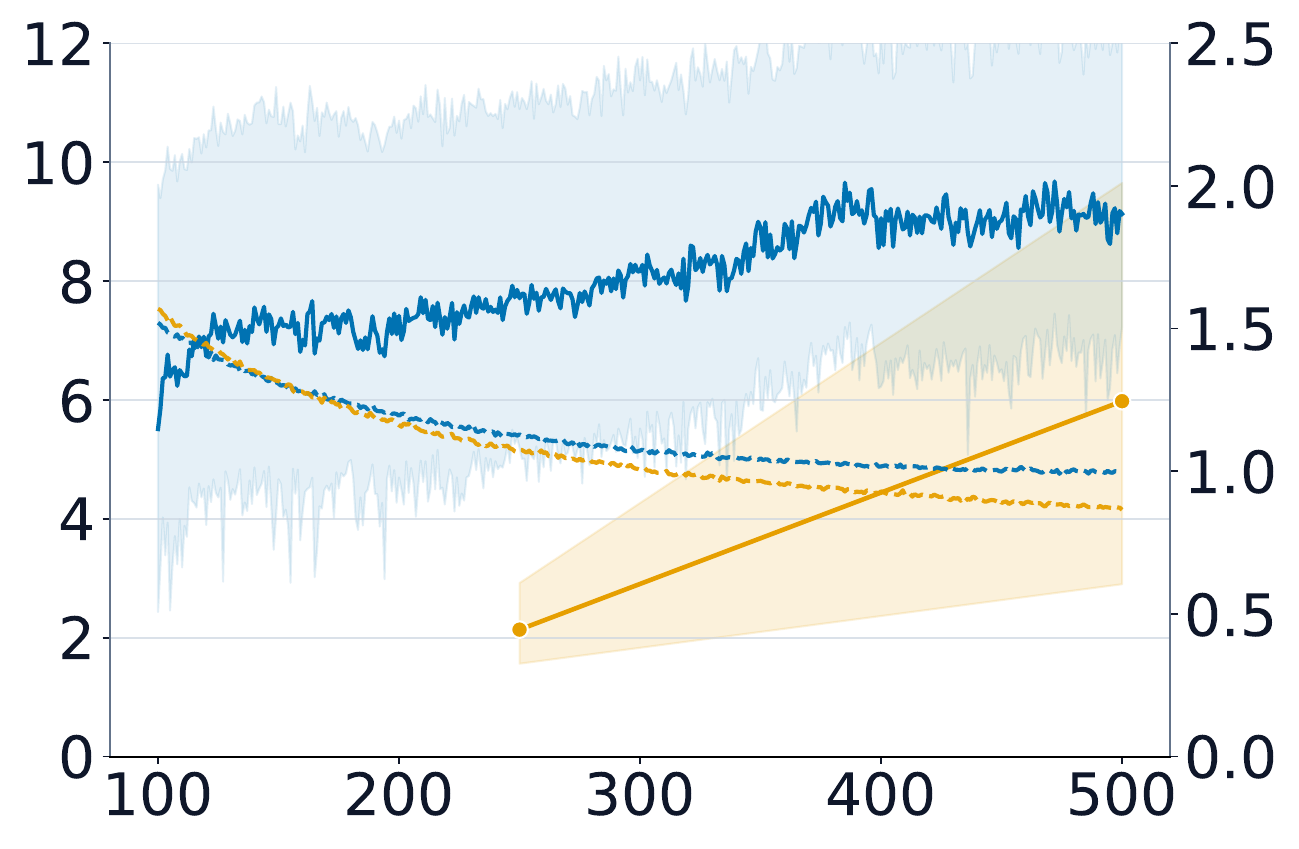}\\[-0.4em]
        {\footnotesize $P=250$}
    \end{minipage}\hspace{0.1cm}
    \begin{minipage}{0.29\linewidth}
        \centering
        \includegraphics[width=\linewidth]{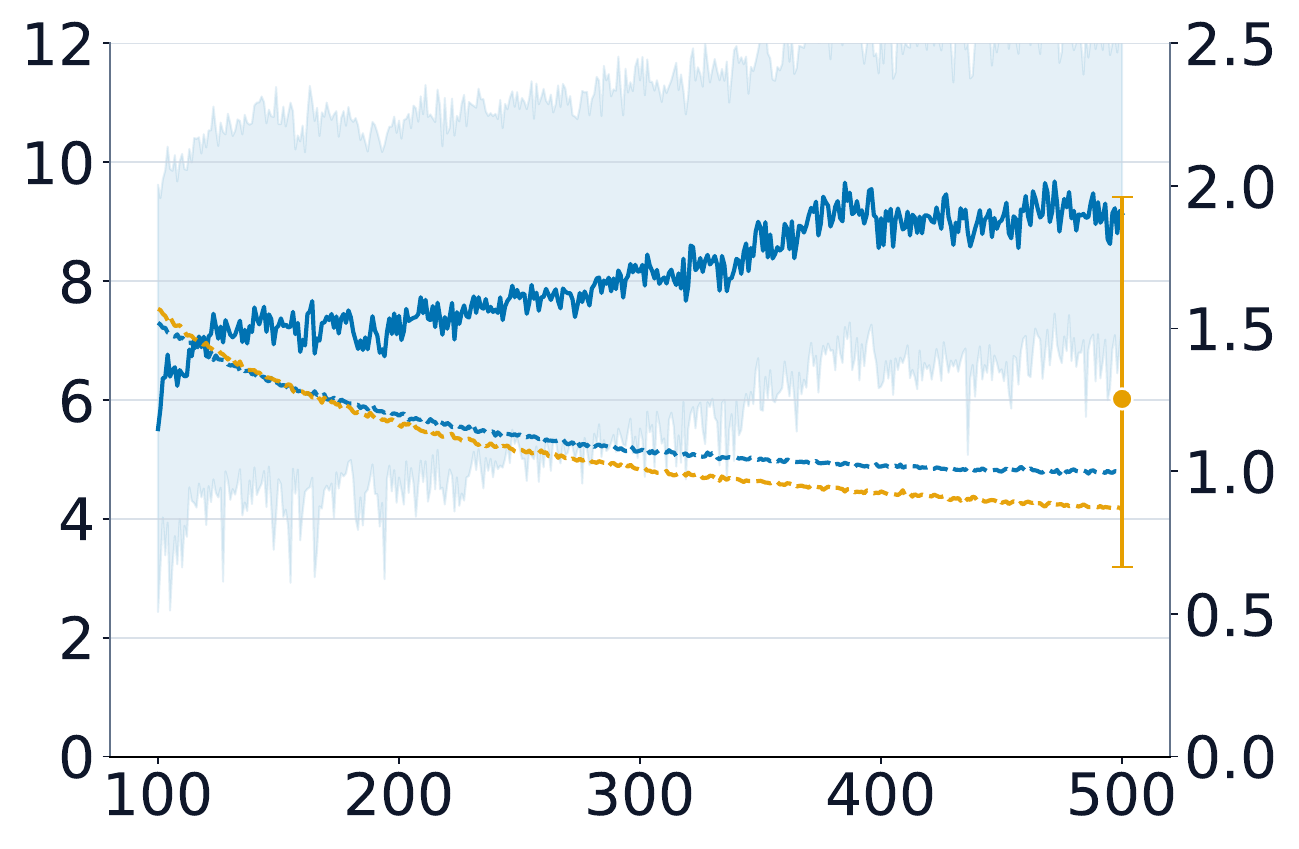}\\[-0.4em]
        {\footnotesize $P=500$}
    \end{minipage}

\end{minipage}
};

\node[rotate=90, anchor=south]
    at ([xshift=0cm]grid.west)
    {Posterior mean of $\epsilon$};

\node[rotate=270, anchor=north, font=\small]
    at ([xshift=-0.45cm]grid.east) {Test loss};

\node[anchor=north]
    at ([xshift=-0.45cm,yshift=-0.2cm]grid.south)
    {Round};
\end{tikzpicture}

\caption{Posterior mean of $\epsilon$ for different inspection periods $P$ ($\sigma=0.02$).
Colors indicate the algorithm: \textcolor{OIblue}{$\mathcal{A}_1$} and
\textcolor{OIorange}{$\mathcal{A}_2$}. Solid lines denote the posterior mean
of $\epsilon$, while dashed lines denote the mean loss. The mean losses of
$\mathcal{A}_2$ and $\mathcal{A}_3$ are identical. Shaded regions indicate
the corresponding 95\% credible intervals.}
\label{fig:Pp_grid}
\end{figure*}

While results in Fig.~\ref{fig:loss_vs_eps} support our claim that the final model can be released privately for downstream use while preserving privacy and accuracy through HE-based training, we also need to evaluate whether intermittent model inspection in meaningful periods can be performed without incurring greater privacy leakage than the DP-only baseline. To this end, Fig.~\ref{fig:Pp_grid} reports the estimated posterior mean of $\epsilon$ across training rounds for different values of $P$, where $P$ denotes the inspection period. For example, when $P=10$, we inspect every 10th model until all $500$ training rounds are complete. 

In our experiments, the results show that even when $P=1$ (i.e., every global model is subjected to inspection), $\mathcal{A}_2$ consistently achieves a lower posterior mean of $\epsilon$ compared to $\mathcal{A}_1$ over the training rounds. At the final round (i.e., round $500$), the estimated privacy loss for $\mathcal{A}_2$ reaches $8.43$, compared to $9.13$ for $\mathcal{A}_1$. This indicates that our framework preserves model utility through HE-based training while still enabling frequent inspections. These results also show that increasing the inspection period improves protocol privacy, which may be preferable for particularly sensitive tasks that do not require as rigorous inspections. It is also important to note that the estimated privacy loss of the protocols is dependent on the variance of noise added, the model architecture, the FL setup, and the properties of the data. Thus, while $\mathcal{A}_2$ incurs less privacy loss than $\mathcal{A}_1$, even for $P=1$ in our setting, this may not necessarily be the case under other experimental configurations.

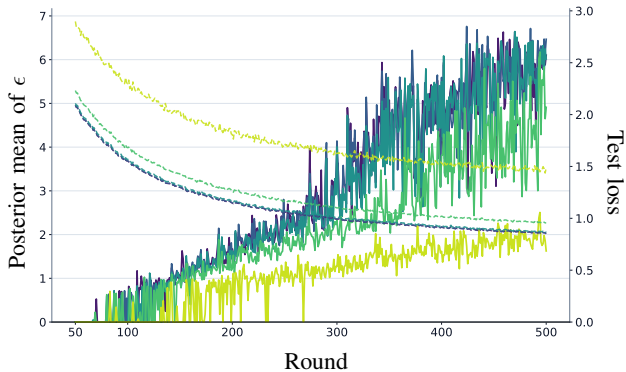
\begin{figure}[!h]
    \centering
    \begin{tikzpicture}
        \node[inner sep=0] (img) {
            \includegraphics[
                width=0.84\linewidth,
                trim={5pt 5pt 5pt 5pt},
                clip
            ]{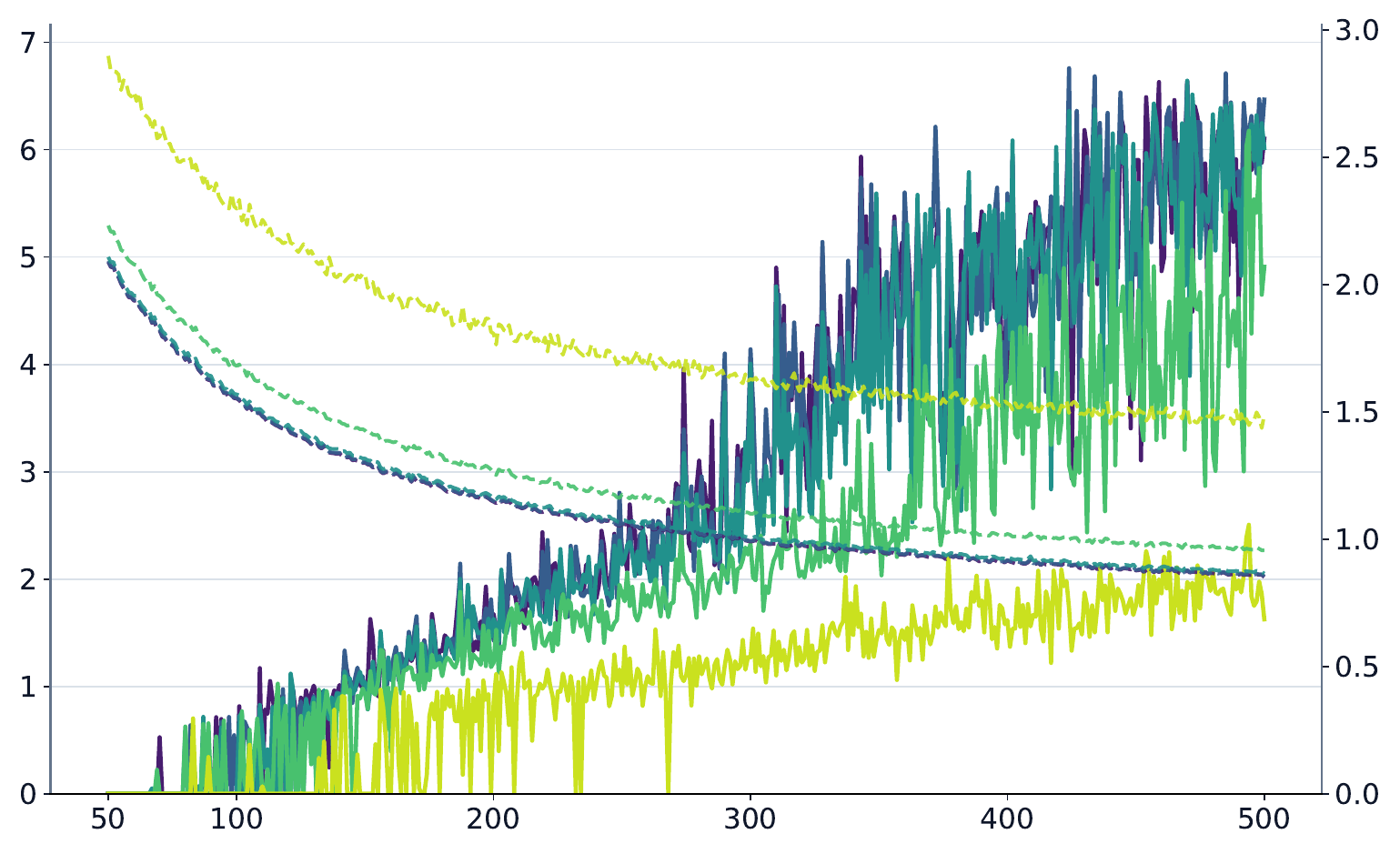}
        };

        \node[rotate=90, anchor=south, font=\small] 
            at (img.west) {Posterior mean of $\epsilon$};

        \node[rotate=270, anchor=south, font=\small] 
            at (img.east) {Test loss};

        \node[anchor=north, font=\small] 
            at (img.south) {Round};
    \end{tikzpicture}

    \caption{Test loss and posterior mean of $\epsilon$ for $\mathcal{A}_3$
    with $\sigma=$
    \textcolor{stdZeroZeroFive}{0.005},
    \textcolor{stdZeroOne}{0.01},
    \textcolor{stdZeroTwo}{0.02},
    \textcolor{stdZeroFive}{0.05}, and
    \textcolor{stdZeroTen}{0.1}.
    Solid and dashed lines denote posterior mean $\epsilon$ and mean test loss,
    respectively.}
    \label{fig:diff_stds}
\end{figure}

Finally, in Fig.~\ref{fig:diff_stds}, we demonstrate the impact of varying Gaussian noise in the $\mathcal{A}_3$ setting. The figure clearly indicates that as noise variance increases, estimated privacy increases, while the model's utility decreases. As expected, this demonstrates that it is possible to achieve significantly more private FL runs if one is willing to sacrifice some accuracy.

\subsubsection{FHE Simulation}
\label{sec:fhe-sim}
In the results presented in Section~\ref{sec:priv-eff-on-acc}, the loss values used to quantify the utility and privacy of $\mathcal{A}_{2/3}$ are generated under the assumption that HE does not affect precision during training. Our privacy estimation procedure requires training $N \times n$ shadow models for each hypothesis, resulting in $4,000$ FL runs in our experimental setting. Training this many models already imposes substantial computational overhead in the cleartext domain. Since homomorphic operations can be orders of magnitude more expensive than cleartext operations, performing fully HE-based training for all shadow models would be computationally impractical for privacy estimation. For instance, FL over a single CNN model using the CIFAR-10 dataset under HE has been reported to take up to a week to complete~\cite{sav_poseidon_2021}. 

Nevertheless, it is essential to validate our assumption and demonstrate that training under HE results in a negligible difference in model utility, thus supporting the claims made in Section~\ref{sec:priv-eff-on-acc}. The difference between cleartext and HE can primarily arise from two factors:
\begin{itemize}[leftmargin=*]
    \item {\em Nonlinear operations}, such as the ReLU and softmax functions used in our model, cannot be applied directly in a homomorphic setting and must be replaced with polynomial or arithmetic approximations.
    \item CKKS is an {\em approximate encryption scheme}. Therefore, a small amount of numerical noise is introduced into the underlying plaintexts during encrypted arithmetic operations.
\end{itemize}

These factors may cause the model to follow a different optimization trajectory, potentially leading to different accuracy and loss values. To ensure that the effect of HE-induced numerical noise and nonlinear approximation remains within an acceptable range, we implemented a simulated FHE version of the FL pipeline for $\mathcal{A}_{2/3}$, using the Alternating Packing (AP) mechanism introduced in \textsc{Poseidon}~\cite{sav_poseidon_2021}. In this implementation, we approximate the ReLU function using the representation $\mathsf{ReLU}(x) = x(1+\operatorname{sgn}(x))/2$, where $\operatorname{sgn}(\cdot)$ denotes the sign function and is polynomially approximated using the composite method in~\cite{he_comparison}. We approximate the softmax function using a Taylor series expansion, Goldschmidt division, and the maximum-finding algorithm presented in~\cite{he_comparison}. 

Moreover, we introduce estimated CKKS numerical noise into each simulated homomorphic operation using Tune Insight's CKKS noise estimation library~\cite{cryptoeprint:2024/853}. This library represents each ciphertext by mock ciphertext components in canonical form and propagates the noisy components through the circuit operation by operation. Furthermore, we adapt a similar noise estimation mechanism for the \texttt{DBootstrap} and \texttt{DBootstrapALT} operations, which are not available in the original library, by comparing the generated noise in our simulated outputs with that produced by the GPU-accelerated FHE library, HEonGPU~\cite{Ozcan2024}.


\begin{figure}[!h]
    \centering
    \begin{tikzpicture}
        \node[inner sep=0] (grid) {
            \includegraphics[
                width=0.86\linewidth,
                trim={5pt 5pt 5pt 5pt},
                clip
            ]{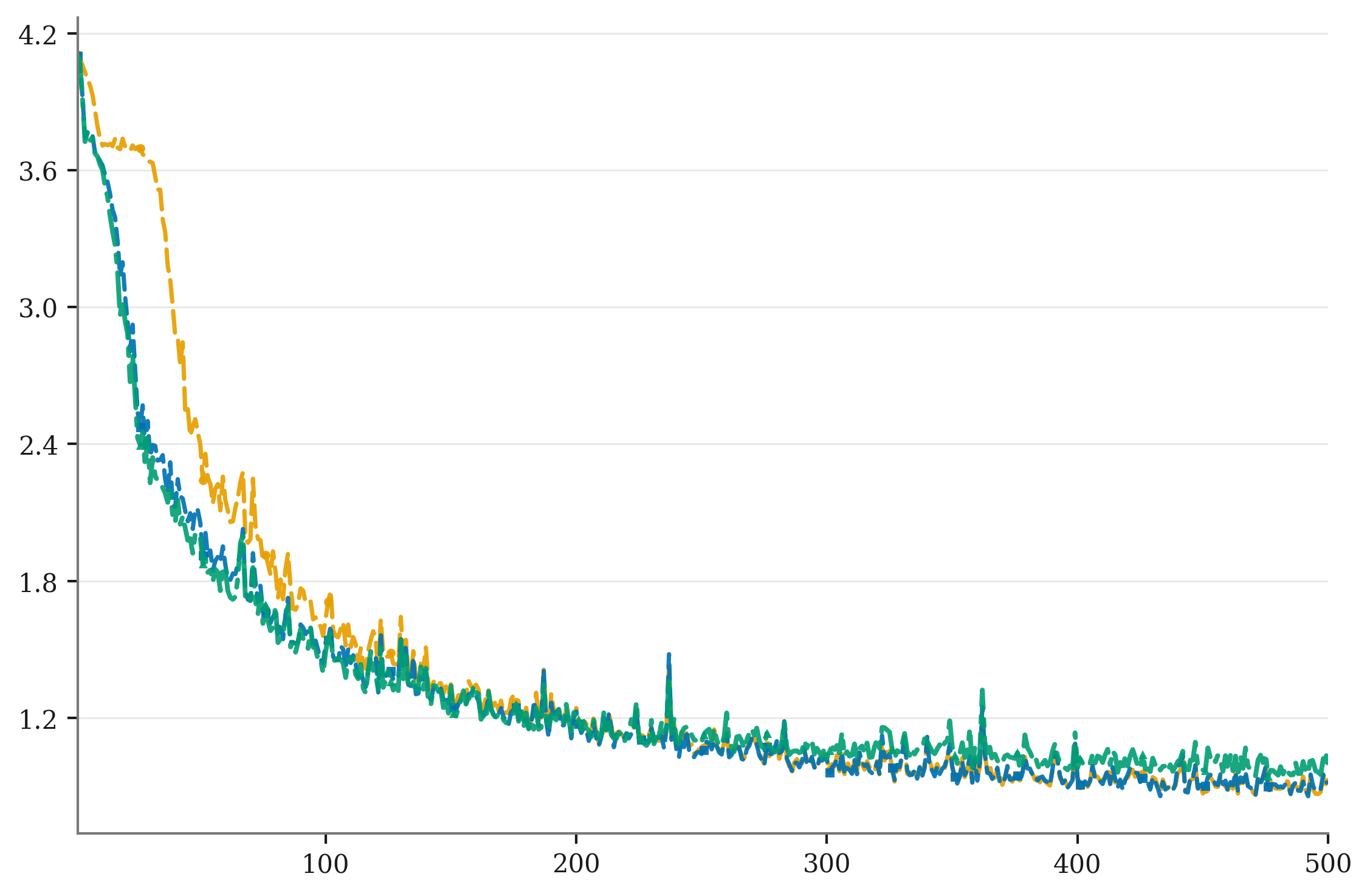}
        };

        \node[rotate=90, font=\small, anchor=south]
            at (grid.west) {Test loss};

        \node[anchor=north, font=\small]
            at (grid.south) {Round};
    \end{tikzpicture}

    \caption{Test loss of \textcolor{OIorange}{FHE-simulated $\mathcal{A}_{2/3}$},
    \textcolor{OIblue}{cleartext $\mathcal{A}_{2/3}$}, \textcolor{OIgreen}{$\mathcal{A}_1$}.}
    \label{fig:he_sim_loss_comp}
\end{figure}

Fig.~\ref{fig:he_sim_loss_comp} compares the test loss of the cleartext alternative of $\mathcal{A}_{2/3}$, utilized in the privacy and utility experiments reported in Fig.~\ref{fig:loss_vs_eps}, Fig.~\ref{fig:Pp_grid}, and Fig.~\ref{fig:diff_stds}, with the FHE-simulated version of $\mathcal{A}_{2/3}$, and the DP-only baseline $\mathcal{A}_1$. The results show that the FHE-simulated version of $\mathcal{A}_{2/3}$ incurs a slight increase in test loss compared with its cleartext counterpart in earlier rounds due to the suboptimal nature of the nonlinear function approximations we use. However, after the model reaches a rudimentary level of convergence, this increase remains negligible and does not reduce the performance of $\mathcal{A}_{2/3}$ below that of $\mathcal{A}_1$. These results further support our results that $\mathcal{A}_{2/3}$ provides comparable test loss to $\mathcal{A}_1$, even when CKKS-induced numerical noise and approximation errors are taken into account.

\section{Conclusion}
\label{sec:conclusion}
In this work, we proposed a utility-preserving FL protocol that enables privacy-preserving model inspection and downstream model release by integrating HE and DP. Our experiments demonstrate that our method offers significantly greater estimated privacy while also providing better utility than the basic approach based solely on DP, which continues training on a previously perturbed model. Consequently, these findings indicate that for use cases requiring higher noise variances to achieve sufficient privacy, our proposed approach can avoid the utility degradation that arises from directly adding DP noise to the model updates used during training. This is because the HE-based approach avoids noise accumulation by only perturbing the final layer of the visible models. Furthermore, we show that model release and intermittent inspections can occur while maintaining estimated privacy levels higher than a strictly DP-based pipeline. These findings are particularly significant because they enable privacy-preserving access to the model during and after training.

As future work, the current HE simulation can be replaced with a full homomorphic implementation to investigate and improve the computational and communication costs of the proposed protocol. In particular, arithmetic and polynomial approximations of the involved nonlinear functions could be optimized to incur less computational overhead in a real-world implementation. Further research could also examine the impact of non-periodic inspections on privacy, as participants may find it more useful to review the model closer to the anticipated end of the training protocol, rather than having frequent inspections at the beginning. Other potential directions include extending our framework to definition (a) by adapting the null and alternative hypotheses accordingly, and exploring various methods for utilizing intermittent models in anomaly detection.

\section*{Acknowledgment}
This work has been supported by the project ``Privacy Compliant Health Data As A Service For AI Development" with grant agreement ID 101095384. This work is funded by the European Union. Views and opinions expressed are, however, those of the author(s) only and do not necessarily reflect those of the European Union. Neither the European Union nor the granting authority can be held responsible for them. 

OpenAI Codex and ChatGPT were used to assist with code generation and debugging for the experimental implementations described in Section~\ref{sec:experimental-results}, including the conversion of the Go-based CKKS noise estimation library~\cite{cryptoeprint:2024/853} into a Python/CuPy-compatible implementation. All resulting code was reviewed and validated by the authors.

\bibliographystyle{IEEEtran}
\bibliography{IEEEtran.bib}

\appendices
\section{Extended Threat Model}
\label{sec:appendix}

In our threat model presented in Section~\ref{sec:threat-model}, we assume that the server cannot collude, even though the clients can. We specifically opted for this assumption for the server, as in large-scale federated learning applications, the servers tend to be operated by large organizations whose behavior and actions are heavily regulated compared to individual, possibly anonymous clients.

Nevertheless, our threat model and protocol can be extended to the setting where the server may collude (with clients). In this section, we define an extended version of our threat model outlined in Section~\ref{sec:threat-model}, which accounts for both the clients and the server colluding as semi-honest entities. We refer to the threat model described in Section~\ref{sec:threat-model} as $\mathrm{T_1}$, and our extended model as $\mathrm{T_2}$.

\subsection{$\mathcal{A}_1$ (DP-only baseline)}

Both the FL algorithm and the adversary's view (collective view of all colluding parties, i.e., the server and $|\mathcal{P}|-1$ clients) are the same as in $\mathrm{T_1}$ (refer to Alg.~\ref{alg:A1} and Eq.~\ref{eq: A1 adv view}, respectively). Therefore, $\mathcal{A}_1$ provides the same privacy guarantees in $\mathrm{T_2}$ as it does in $\mathrm{T_1}$. Its privacy can be estimated using the attack definition outlined in Section~\ref{sec:miatfdp}. The only distinction is that, since the server is responsible for selecting clients each round for training ($\mathcal{C}^{(r)} \overset{{\scriptscriptstyle\$}}{\leftarrow} \mathcal{T}, \forall r$) and informing them, the collective can now possess the knowledge of the honest clients' identities directly. This removes the standard membership inference implications associated with our MIA in Definition~\ref{defn: MIAtfDP}.

\subsection{$\mathcal{A}_{2/3}$ (HE-based training)}

In the case of $\mathcal{A}_{2/3}$, Alg.~\ref{alg:A2} does not maintain the same level of privacy under the colluding server assumption. In a given inspection round $r$, because the server adds the noise term $V^{(r)}$ to the global model update ${\theta}^{(r)}$, the colluding coalition can simply subtract $V^{(r)}$ from the released model and recover the unnoised global update. Then, with the $|\mathcal{P}|-1$ colluding clients' local updates, they can reconstruct the honest client’s update ${\theta}_{c_h}^{(r)}$ with no noise. Thus, this algorithm definition does not protect $c_h$'s sensitive data against adversaries.

To address this issue, clients can collectively generate the noise to be added. Each client $c \in \mathcal{C}^{(r)}$ can generate its own noise term, encrypt it, and send it to the server for it to be added to the inspection model. In this way, the semi-honest server cannot see the noise terms of the clients, but adds each term to the model update as it follows the protocol steps correctly. However, this is also not an appropriate solution. If a noisy model is opened for inspection in a round where $c_h \notin \mathcal{C}^{(r)}$, the colluding clients can remove their noise terms from the update and reveal the non-noisy model ${\theta}^{(r)}$, which may indirectly carry information about $c_h$'s data, as $c_h$ might have participated in the training in earlier rounds.

As a result, we want a mechanism that protects $c_h$'s contribution (with some noise) even during inspection rounds in which $c_h$ is not participating. To ensure this, we propose the following scheme. 

In our proposed protocol, we update the \emph{client collective noise generation} idea to include all clients $c\in \mathcal{T}$, regardless of whether they participate in the current round $r$. If each client $c \in \mathcal{T}$ sends an encrypted Gaussian noise term $[W_{c}]$ to the server, the server can then carry out the aggregation as
\[
[\tilde{\theta}^{(r)}] = \frac{\sum_{c \in \mathcal{C}_{r}} n_{c} [\theta_{c}^{(r)}]}{\sum_{c \in \mathcal{C}_{r}} n_{c}} + \sum_{c \in \mathcal{T}} [W_{c}].
\]

With everyone's cooperation, $[\tilde{\theta}^{(r)}]$ is collectively decrypted by the clients and $\tilde{\theta}^{(r)}$ is published for inspection. This formulation thus ensures that $c_h$'s noise remains present in the shared model update, even if all $|\mathcal{P}|-1$ clients and the server collude. 

To ensure that the level of noise added to the published model matches $\mathcal{A}_1$, i.e., to have a variance of  $\sum_{c\in \mathcal{C}^{(r)}}\alpha_c^2\sigma^2$, the client-level noise should be defined as
\[
W_{c} \sim \mathcal{N}(0, \beta^2\sigma^{2} I_d)
\]
provided that
\[
\sigma^2\sum_{c\in\mathcal{T}}\beta^2
= 
\sigma^2\sum_{c\in \mathcal{C}^{(r)}}\alpha_c^2,
\quad 
\beta = \sqrt{\frac{\sum_{c'\in \mathcal{C}^{(r)}}\alpha^2_{c'}}{{K}}}.
\]
This setup provides less privacy per inspection round than the non-colluding server assumption. This is because the colluding clients can remove the noise from the inspected model and view the model as 
\[
\mathsf{View}'^{(R)}_{\mathcal{A}_{2/3}} =  \left\{\sum_{c \in \mathcal{C}^{(r)}}\alpha_c\theta_{c}^{(r)} + W_{c_h} : r \in \mathcal{I}^{(R)}_{\mathcal{A}_{2/3}} \right\}.
\]
At the individual level, contribution from the honest client $c_h$ is effectively protected with noise ${W_{c_h}}/{\alpha_{c_h}}$ (note that $\alpha_{c_h} <1$), which has a variance of $\beta^2 \sigma^2 / \alpha_{c_h}^2$. Although $\theta_{c_h}^{(r)}$ cannot be revealed, even in a coalition of $|\mathcal{P}|-1$ colluding clients and a colluding server, this analysis shows that $c_h$'s contribution is subject to less noise.

The relationship between the noise differences becomes clearer if we assume that the sizes of the clients' datasets, $n_c : c\in\mathcal{T}$, are equal. In that case, the client noise can be represented as 
\[
W_{c} \sim \mathcal{N}(0, \sigma^{2} I_d/(K_{s}K)).
\]
After the colluding clients remove their noise, the adversary's view becomes
\[
\mathsf{View}'^{(R)}_{\mathcal{A}_{2/3}} = \left\{
\frac{1}{K_{s}} \sum_{c \in \mathcal{C}_{r}}\theta_{c}^{(r)} + W_{c_h} : r \in \mathcal{I}^{(R)}_{\mathcal{A}_{2/3}}\right\},
\]
where $c_h$'s contribution is effectively protected with noise $K_{s} W_{c_h}$, which has a variance of $K_{s}\sigma^{2}/K$.

\end{document}